\documentclass[12pt]{article}
\usepackage{epsfig}
\newcommand{\dubbelR}{{\sf I}\kern-.12em{\sf R}}
\newcommand{\dubbelN}{{\sf I}\kern-.12em{\sf N}}
\newcommand{\dubbelZ}{\mbox{\sf Z\hspace*{-5.4pt}Z}}
\newfont{\Bbb}{msbm10 scaled\magstep1}
\newcommand{\CC}{\Bbb{C}}

\def\unit{{\sf1 \! I}}
\def\half{{\mbox{\scriptsize{$\frac{1}{2}$}}}}
\newcommand{\ij}{i\kern -0.08em j}

\newcommand{\be}{\begin{equation}}
\newcommand{\ee}{\end{equation}}
\newcommand{\bea}{\begin{eqnarray}}
\newcommand{\eea}{\end{eqnarray}} 
\newcommand{\nn}{\nonumber}
\newcommand{\dubbelint}{\mathop{\int\int}\limits_{\kern-5.5pt {V-V_1}}}
\newcommand{\al}{\alpha}
\newcommand{\bt}{\beta}
\newcommand{\gm}{\gamma}
\newcommand{\Gm}{\Gamma}
\newcommand{\dl}{\delta}
\newcommand{\Dl}{\Delta}
\newcommand{\ep}{\epsilon}
\newcommand{\varep}{\varepsilon}

\newcommand{\tht}{\theta}

\newcommand{\kp}{\kappa}
\newcommand{\lm}{\lambda}
\newcommand{\Lm}{\Lambda}

\newcommand{\sg}{\sigma}
\newcommand{\Sg}{\Sigma}

\newcommand{\om}{\omega}

\newcommand{\winv}{w^{-1}}
\newcommand{\xinv}{x^{-1}}
\newcommand{\yinv}{y^{-1}}
\newcommand{\zinv}{z^{-1}}
\newcommand{\ginv}{g^{-1}}
\newcommand{\hinv}{h^{-1}}

\newcommand{\fog}{f \otimes g}
\newcommand{\aphi}{{}^{A}\!\phi}
\newcommand{\bphi}{{}^{B}\!\phi}
\newcommand{\rphi}{{}^{r}\!\phi}
\newcommand{\ophi}{{}^{0}\!\phi}

\newcommand{\ABPhi}{{}^{A,B}\!\Phi} 
\newcommand{\rrPhi}{{}^{r_1,r_2}\!\Phi_{n_1,n_2}} 
\newcommand{\rD}{{}^{r}\!D} 
\newcommand{\reD}{{}^{r_1}\!D} 
\newcommand{\rtD}{{}^{r_2}\!D} 
\newcommand{\cA}{{\cal A}}
\newcommand{\DA}{{\cal D}({\cal A})}
\newcommand{\DG}{{\cal D}(G)}
\newcommand{\DS}{{\cal D}(SU(2))}

\newcommand{\Nabc}{N^{AB\gamma}_{\al\beta C}}
\newcommand{\cJ}{{\cal J}}
\newcommand{\cP}{{\cal P}}
\newcommand{\emu}{e_{\mu}}
\newcommand{\omu}{\om_{\mu}}
\newcommand{\De}{D^{j_1}_{m_1 n_1}} 
\newcommand{\rDe}{{}^{r_1}\!D^{j_1}_{m_1 n_1}} 
 
\newcommand{\Dt}{D^{j_2}_{m_2 n_2}} 
 
\newcommand{\Dd}{D^{j_3}_{m_3 n_3}}

\newcommand{\Conj}{{\rm Conj}}
\newcommand{\iso}{ISO(3)}

\newcommand{\D}{{\cal D}}
\newcommand{\va}{\vec{a}}
\newcommand{\vj}{\vec{j}}
\newcommand{\vep}{\vec{p}}

\input{amssym.def}
\input{amssym.tex}
\def\gog{{\goth g}}
\def\gosu{su(2)}
\def\goiso{isu(2)}
\newcommand{\semdir}{\ltimes}

\newcounter{fignum}

\newtheorem{thm}{Theorem}[section]

\newtheorem{defn}[thm]{Definition}
\begin{document}
\title{
\hfill{\normalsize UvA-WINS-ITFA 98-07}\\[-0.3 cm]
\hfill{\normalsize hep-th/9804130}\\[1.5 cm]
   Topological field theory and \\the quantum double of
   $SU(2)$} 
\author{F.A.\ Bais\thanks{email: bais@phys.uva.nl}  and 
       N.M.\ Muller\thanks{email: nmuller@phys.uva.nl}
     \\Institute for Theoretical Physics, University of Amsterdam, \\
       Valckenierstraat 65,
       1018 XE Amsterdam, The Netherlands.}
\maketitle

\begin{abstract}
We study the quantum mechanics of a system of topologically
interacting particles in 2+1 dimensions, which is described by
coupling the particles to a Chern--Simons gauge field of an
inhomogeneous group. Analysis of the phase space shows that for the
particular case of $\iso$ Chern--Simons theory the underlying symmetry
is that of the quantum double $\D(SU(2))$, based on the homogeneous
part of the gauge group. This in contrast 
to the usual $q$-deformed gauge group itself, which occurs in the case
of a homogeneous gauge group. Subsequently, we describe the
structure of the quantum double of a continuous group and
the classification of its unitary irreducible representations. The
comultiplication and the $R$-element 
of the quantum double allow for a natural description of the fusion 
properties and the nonabelian braid statistics of the particles. These
typically manifest themselves in generalised Aharonov--Bohm scattering
processes, for which we compute the differential cross sections.
Finally, we briefly describe the structure of ${\cal D}(SO(2,1))$, the
underlying quantum double symmetry of (2+1)-dimensional quantum gravity. 
\end{abstract}
PACS: 03.65.Fd ; 11.15.-q ; 11.25.Hf\\
Keywords: topological interactions; Chern--Simons theory; quantum
double groups

\section{Introduction}
\setcounter{equation}{0}
The study of topological quantum field theories has lead to important
results in both mathematics and physics \cite{Danny}. For example, the
intrinsically three dimensional 
formulation of knot and link invariants is given by correlation
functions in three dimensional Chern--Simons (CS) theory. Also, it is known
that the Hilbert space of such Chern--Simons theories correspond to
the chiral sectors of certain conformal field theories
\cite{WitJonPol} \cite{EHVer}, and that the fusion and braid
properties of these sectors can be described via so-called `hidden
quantum symmetries', see \cite{GRS} and references therein. More
specifically, the wave functionals in the Wess--Zumino--Witten (WZW)
model for which the chiral algebra is the Kac--Moody algebra
$\widehat{SU(2)}_k$ obey the
braiding and fusion relations of the quantum group $U_q(su(2))$, where
the deformation parameter $q$ is related to the level $k$ of the WZW
model, which in turn is the coupling constant of the corresponding CS
theory. These models have found interesting applications in physical
systems as for example the quantum Hall effect, Kondo-like impurity
problems and high-$T_c$ superconductivity, and are also directly
linked to the theory of integrable two-dimensional models in
statistical mechanics. 

In this paper we are concerned with the quantum description of
particles which carry charges of the inhomogeneous group
$\iso$, and which interact topologically through their coupling to a
CS gauge field. So one should think of particles which are
characterised by some internal ``Euclidean mass'' $r$ and a spin $s$. 
To be precise, we actually consider the double cover of $\iso$, which is
the semi-direct product of $SU(2)$ with the group of translations in
three dimensions, but for convenience we denote the gauge group by $\iso$. 
It turns out that due to the
inhomogeneous structure of the gauge group the physical phase space of the
system has a natural cotangent bundle structure, which directly leads
to the identification of the  configuration space. We show that the
internal degrees of freedom  correspond to holonomies $g\in\iso$
of noncontractable loops in physical space, which leads to a
noncommutative multiplicative structure of the multi-particle
configuration space. 
%Secondly, there are degrees of freedom in the
%representation spaces of irreducible representations of 
%the centraliser subgroups $N_g$ which leave the holonomies invariant.
Upon quantisation of the single particle sectors one obtains that
states in our system carry a `magnetic flux' (topological charge)
which is the holonomy $g \in SU(2)$, and an `electric charge' $n$
which labels some irreducible representation of the centraliser
subgroup $N_g \subset SU(2)$. These charges are not identified via Gauss'
constraint, in contrast to the case of `ordinary' $SU(2)$
Chern--Simons theory, where the phase space does not have a natural
cotangent bundle structure.  
The multi-particle sectors of the theory should exhibit two important
features: firstly, the fusion properties have to be defined on the  
level of the multi-particle Hilbert space. Secondly, 
the topological interactions between the particles should be
described by a suitable action of the braid group on this
multi-particle Hilbert space. We show that for the model we are
considering the correct mathematical structure, leading to the required
braiding and fusion properties, is the so-called quantum double $\DG$
of the {\sl homogeneous} part of the gauge group, in this case $G=SU(2)$. 
This implies that the internal Hilbert space can be decomposed into irreducible
unitary representations of this quantum double ${\cal D}(G)$, and we
will show that the labels of these representations precisely coincide
with invariants associated to the `magnetic' and `electric' charges
mentioned above, i.e.\ the internal mass and spin.    

This $\iso$ Chern--Simons theory may be referred to as Euclidean
quantum gravity in 2+1 dimensions. To a certain extent we do indeed
solve this problem, in the sense that we explicitly describe the
Hilbert space and the underlying quantum symmetry. At the end of the
paper we also do this for the realistic, Minkowskian version, i.e.\
the group $ISO(2,1)$. However, we do not want to over-emphasize the
gravity aspect of our work, because in this paper we do not address
the problem of identifying the internal ``Chern--Simons'' degrees of
freedom with their external space time counterparts. We do return to
this issue in a forthcoming paper \cite{BMgrav}

At this point it may be interesting to note that quantum doubles have
featured before in physics. For example, in 
\cite{DPR} it has been shown that the sectors in certain rational
conformal field theories (orbifold models) correspond to the
irreducible representations of $\D(H)$, where $H$ denotes a finite group.
The same quantum doubles have been used to describe the topological
interactions of defects carrying discrete magnetic fluxes in gauge
theories which are spontaneously broken to a finite residual symmetry
group \cite{BWP} \cite{WPB}. It was observed that quantum states of such
systems can be exactly described by elements of representation spaces
of the quantum double of the residual (finite) group. As in our case,
the use of this observation lies in the fact that the quantum double
as underlying symmetry for these systems automatically comprises the
correct notions of `fusion', via the decomposition of tensor product
representations, and `braiding', via the action of the $R$-element of
the quantum double. The identification of this quantum double symmetry
allowed for explicit computation of differential cross sections in
nonabelian generalisations of the Aharonov--Bohm effect. Finally, 
the associated $S$-matrix diagonalising the fusion rules generates an
electric--magnetic duality transformation in these models. 
Both models just mentioned are related to the discrete topological
field theories of Dijkgraaf and Witten \cite{DW}. In the present paper
we consider a more intricate class of models which exhibit a
manifest {\sl continuous} quantum double symmetry. 
\vspace*{.2cm}\\
The outline of the paper is as follows:\\
We start by describing the physical system of particles
that carry quantum numbers of the group $\iso$, and that have topological
interactions mediated by an $\iso$ Chern--Simons gauge field. In
section \ref{s2} we consider the pure $\iso$ Chern--Simons
theory, without particles, and construct its physical phase
space. Subsequently, in section \ref{s2a} we include point particles,
and arrive at an explicit parametrisation of the corresponding phase
space. In section \ref{squant} we describe the quantisation of the single
particle case. Next we discuss the nontrivial properties of the
multi-particle 
quantum system, involving the aformentioned `braiding' and 
`fusion'. Section \ref{s3} summarises the mathematical results of
\cite{KM} and \cite{KBM} on the quantum double of a (locally) compact
group. In section \ref{s4} we turn to the quantum double of $SU(2)$
explicitly. We identify the labels of the irreducible unitary
representations of $\D(SU(2))$ as the quantum numbers we discussed
before, and show that elements in the corresponding representation
spaces precisely coincide with the states we found by quantising the
single particle sectors. Also, it is shown that the tensor products of
such irreducible representations possess precisely the correct
decomposition and braiding properties required for the quantised
multi-particle sectors. In section \ref{s5} we give the detailed
formulae for the fusion rules, which are the multiplicities of the 
Clebsch--Gordan {\sl series} in $\DS$, for all types of product
representations. We also compute the Clebsch--Gordan {\sl
coefficients}. One important result is that we
can calculate differential cross--sections for various generalised
Aharonov--Bohm scattering processes, because we know the 
$R$-element and its action on tensor product states, this is discussed
in section \ref{s6}. In section \ref{sgrav} we show that there is a
quantum double structure in 2+1 quantum gravity, corresponding to ${\cal
D}(SO(2,1))$, and in section \ref{s7} we present our conclusions.
\section{The pure $\iso$ Chern--Simons theory}\label{s2}
\setcounter{equation}{0}
We start by considering pure Chern--Simons theory in 2+1 dimensional
spacetime $M$, because this will reveal much about the general
structure of our classical phase space. It is convenient to include
the particles later. For an introduction to CS theory we refer to
\cite{Danny} and references therein. The action is given by  
\be
S_{CS} = \frac{1}{2} \int_M \mbox{tr}(A\wedge dA + \frac{2}{3}A\wedge
A\wedge A) 
\label{eq:CSaction}
\ee
for $M$ an --in principle arbitrary-- oriented three dimensional
manifold, and a gauge field $A$ taking values in the Lie algebra
$\gog$ of the symmetry group $G$. This defines a topological theory 
because the Lagrangian, which is a 3-form without metric dependence,
is integrated over a three dimensional manifold, so the action is
independent of the metric on $M$.    

The case of compact $G$ has been well studied, and on the quantum
level it is related to the $q$-deformation of $G$, or 
better, to the $q$-deformation of the universal enveloping algebra of
$\gog$, see for instance \cite{GRS}. In the present paper we consider a
particular example of a noncompact, inhomogeneous group, being the
semi-direct product of the groups of rotations and translations in
three (internal) dimensions with Euclidean signature. 
The commutation relations read
\be
[\cJ_a, \cJ_b]=\ep_{abc} \cJ^c,\quad [\cJ_a,\cP_b] = \ep_{abc}\cP^c,
\qquad [\cP_a,\cP_b] = 0,
\label{eq:comreliso3}
\ee
with $\cP_a$ the generators of translations and $\cJ_a$ the generators
of $SO(3)$, and together they span the algebra of the (double cover of
the) Euclidean group
in three dimensions: $E(3) = ISO(3)$. The trace in Eq.(\ref{eq:CSaction})
stands for the inner product on the algebra, which we take to be  
\be
\langle\cJ_a, \cP_b\rangle = \dl_{ab},\quad 
\langle\cP_a, \cP_b\rangle = \langle\cJ_a, \cJ_b\rangle = 0.
\label{eq:innerproduct}
\ee
The fact that the inner product is not unique for this case was 
noted in \cite{Witgrav}.

The $ISO(3)$ gauge field is given by
\be
A_{\mu}(x) = \emu^a (x)\cP_a + \omu^a(x)\cJ_a,\quad x\in M,\; a=0,1,2.
\label{eq:afield}
\ee
The components $\emu^a$ and $\omu^a$ of the gauge field are taken to
be independent variables.
The fact that we are dealing with an inhomogeneous gauge group has the
important consequence that we can scale away any coupling constant in
front of the action, simply by absorbing it in the inhomogeneous part
of the gauge field. In the case of homogenous $G$ the coupling
constant $k$ cannot be scaled away, and it will label physically
distinct quantum theories via the deformation parameter $q$. For
example, for the $SU(2)$ WZW-model at level $k$ we have that $q =
e^{\frac{i\pi}{k+2}}$. The fact that in the $iso$ CS model we considera
the coupling constant can be scaled away, indicates that no
`ordinary' $q$-deformed quantum group will show up.

Observables in this theory must be topological invariants, so independent
of the metric. As argued by Witten \cite{WitJonPol} these correspond
to  Wilson lines,
defined to be the following functionals of the gauge field
\be
W_R(C) := \mbox{tr}_R P \exp \oint_C A.
\label{eq:loop}
\ee
Here $C$ denotes a closed loop in $M$, and the trace is taken in some
irreducible representation $R$ of the gauge group $G$. Later on we
will orient the loop $C$ in the direction in $M$ that we call `time',
and consider some region in $M$, where we may interpret these Wilson
lines as the world lines of the particles. 

Finally, we remark that this (pure) $ISO(3)$ CS theory is completely
equivalent to a (pure) $SU(2)$ BF-theory on a three-dimensional
closed manifold $M$. The latter are topological gauge
theories, which for the case of nonabelian groups have the action
\be
S_{BF} = \int_{M} \mbox{tr} B\,F.
\ee
Here $B$ is a Lie algebra valued 1-form, and $F$ is the curvature of some
flat $G$-bundle over $M$. More details can be found in \cite{Danny}. 

\subsection{Phase space}\label{s22}
The classical field equations from the Chern--Simons
Lagrangian constrain the curvature $F$ of the connection $A$ to be
zero. This means that the solutions are flat connections, 
\be
F_A = 0.
\label{eq:f0}
\ee
In order to canonically quantise the theory we have to
identify the phase space, which requires splitting of space and time
coordinates in the action. As usual we take $M=\Sg\times \dubbelR$, so
two-dimensional Riemann surfaces $\Sg$ as cross-sections of $M$ that
move in (uniform) time. For later convenience we parametrise the
surface $\Sg$ by $z$. The action now reads
\be
S = \int dt \int_{\Sg} d^2 z\left(-2\varep^{ij} e^a_i \partial_t \om_{ja}
 + e^a_0 R_a + \om^a_0 T_a\right)
\label{eq:splitact}
\ee
where we have defined the components of the curvature $F(z) = R_a(z) \cJ^a 
+ T_a(z) \cP^a$ as
\bea
R_a &=& \half\varep^{ij} (\partial_i\om_{ja} -\partial_j\om_{ia} +
\ep_{abc}\om^b_i\om^c_j),\nn\\
\label{eq:ra}
T_a &=& \half\varep^{ij}(\partial_i e_{ja} -\partial_j e_{ia} +
\ep_{abc}(\om^b_i e^c_j - \om^c_j e^b_i)).
\label{eq:ta}
\eea
The dynamical variables are $e^a_i$ and $\om^a_i$, and their Poisson
brackets which can be derived from the action read
\bea
\{ \om^a_i(z_1), e^b_j(z_2)\} &=& \half\dl_{ab}\, \varep_{ij}\,
 \dl^{(2)}(z_1-z_2)\nn\\
\{ e^a_i(z_1), e^b_j(z_2)\} &=&\{ \om^a_i(z_1), \om^b_j(z_2)\} = 0.
\eea
In the action $e^a_0$ and $\om^a_0$ appear as Lagrange multipliers, 
variation with respect to them yields the constraints
\be
R_a = 0,\qquad T_a = 0, 
\label{eq:fis0}
\ee
with Poisson brackets
\bea
\{ T_a(z_1), T_b(z_2)\} &=& \ep_{abc} T^c(z_1) \dl^{(2)}(z_1-z_2),\nn\\
\{ T_a(z_1), R_b(z_2)\} &=& \ep_{abc} R^c(z_1) \dl^{(2)}(z_1-z_2),\nn\\
\{ R_a(z_1), R_b(z_2)\} &=& 0.
\label{eq:RTbracks}
\eea

In general, constraints can either be imposed before, or after
quantisation. If the unconstrained system is quantised, then the
constraints have to be imposed as operators annihilating the physical
states. However, for the current problem of quantisation of a CS gauge
field, it is 
more useful to impose the constraints on a classical level, because the
resulting reduced phase space --called the physical phase space--
turns out to be finite dimensional. The unreduced phase space is infinite
dimensional, it consists of all gauge connections on $\Sg$. The physical
phase space is the space of solutions to the equations of motion which
satisfy the constraints, divided out by the gauge group ${\cal G}$ generated 
by the constraints. So in our case it is the moduli space of flat $\iso$
connections on the Riemann surface $\Sg$:  
\be
{\cal M}_{\mbox{\scriptsize{phys}}} = \{\mbox{flat $\iso$ connections
on}\;\Sg\}/ {\cal G}.
\label{eq:phsptot}
\ee
This is not a very useful description if one wants to perform
(canonical) quantisation of the system, and we recall the following
argument (see also \cite{Car, Danny}) in order to obtain a more explicit
and practical description. A flat connection on a surface is, up to a
gauge transformation, completely determined by its holonomies around
noncontractable loops with a fixed base point on the surface. Such
holonomies only depend on the homotopy class of the loops, not on the
precise paths, so they are homomorphisms from the
(based) fundamental group of the surface to the relevant symmetry group,
here $\iso$. Gauge transformations only have an effect at the base point $*$,
where they act on the holonomies by conjugation with ${\cal G}(*)$.
Hence one arrives at 
\be
{\cal M}_{\mbox{\scriptsize{phys}}} = \mbox{Hom}(\pi_1(\Sg,*);\iso)/\sim
\label{eq:phsphom}
\ee
where $\sim$ denotes equivalence under conjugation by ${\cal
G}(*)$. This expression for the physical phase space shows that if
there are no noncontractable loops in $\Sg$ the theory is trivial. In
the next section we consider the case where noncontractable loops are
present.

Finally, how is a classical state, being an element in ${\cal
M}_{\mbox{\scriptsize{phys}}}$, characterised? A point in the space of
homomorphisms is described by the collection of all $*$-based
noncontractable loops (more precisely, their homotopy classes) in $\Sg$,
together with the values of their corresponding holonomies. It suffices to
give the holonomies corresponding to a certain set of basis elements of
$\pi_1(\Sg,*)$, because they determine the values of the holonomies
corresponding to all other noncontractable loops. The gauge group 
acts via simultaneous conjugation with an element of ${\cal G}(*)$ of all
(basis) holonomies, and has to be divided out. 

\section{Including point particles on the classical level}\label{s2a}
From now on we assume that the spatial part $\Sg$ of our 2+1
dimensional space time has no handles. To make the theory
nontrivial we allow for punctures in $\Sg$, and at these points
insert particles carrying $\iso$ charges. In $M$ these punctures sweep out 
world lines, which will form a certain braid. Since the particles
carry nonabelian charges, one may expect that on the quantum level the braid
group will have a nontrivial realisation on the multi-particle Hilbert
space, which may give rise to a (nonabelian) generalisation of the
Aharonov--Bohm effect. We return to this issue in section \ref{s6}.

Point particles can be introduced in various ways. In what is usually called
a `dynamical' way, the Riemann surface $\Sg$ is considered to have a
boundary which consists of infinitesimally small circles with their interior
removed, and a vanishing curvature everywhere on $\Sg$. This way the degrees
of freedom are contained in the values of the gauge field on the boundary,
and in the matching conditions of the simply connected patches with
vanishing curvature that build up the non-simply connected surface $\Sg$.
This is the approach Witten suggested at the very end of \cite{WitJonPol}.
By taking the three manifold $M$ the way we have described it
above, the space of conformal blocks of a certain conformally
invariant theory in 1+1 dimensions will be equivalent to the Hilbert space 
of a canonically quantised Chern--Simons model with charged point
particles in 2+1 dimensions \cite{EHVer}.
 
For multi-particle configurations this
description via boundary conditions can get rather complicated, and in this
paper we take a different route. We introduce particles `by hand', by
considering the removed points to actually be part of $\Sg$, in such a way
that the gauge field (and its curvature) can take values at these points,
but loops cannot be pulled through them, i.e.\ these points still give rise
to noncontractable loops. At the punctures we allow the gauge field to 
have curvature, so the constraints of Eq.(\ref{eq:fis0}) get modified in the
presence of charged particles. They receive delta function contributions on
the right hand side, and we will now explain how this can be done in a
consistent way for a single particle. 

\subsection{The single particle case} \label{sssingleclas}

We have seen that the observables in the theory come from Wilson lines as
given in Eq.(\ref{eq:loop}), and that they are directly related to
irreducible representations of the gauge group. Following Witten
\cite{WitJonPol}, we consider a timelike Wilson line, i.e.\ the closed loop
$C$ in Eq.(\ref{eq:loop}) running along the time direction and closing at
infinity, to be the world line of a particle. The charges of the particle
correspond to the irreducible representation $R$ associated with the Wilson
line. Such a Wilson line will intersect a spacial slice $\Sg$ at a certain
time $t$ in the point $z_1\in \Sg$, where it is considered as a nonabelian
charge acting as a source for the curvature. 

As pointed out above Eq.(\ref{eq:phsptot}) we want to arrive at the modified
constraint equations (with sources) on a classical level. Therefore, we must
find the classical equivalent of the charges of the particle, which on the
quantum level reproduce the generators of the gauge symmetry at the position of
the particle. To that aim we
introduce a suitable classical mechanical system, which we associate to the
Wilson line, such that upon quantisation we obtain the irreducible
representation $R$. As is known from representation theory \cite{Kir} an
appropriate classical system is one which has a certain coadjoint
orbit of the gauge group as its classical, reduced phase space.
Subsequently, this classical system has to be
coupled to the CS gauge field, effectively making the nonabelian charges
interact topologically. This way one is indeed led to constraint
equations where the symmetry generators of the classical system show
up as the sources for the curvature. In the following subsections we
will explain this procedure in more detail. 

\subsection*{Coadjoint orbit method}
For semi-simple Lie groups the coadjoint orbit method of Kirillov and
Kostant allows one to construct the action for a classical system in  
such a way that the generators corresponding to the Noether symmetry
under Poisson brackets satisfy the algebra
isomorphic to the Lie algebra 
\cite{Kir}. Quantisation will then automatically lead to an irreducible
unitary representation of the corresponding Lie group. In fact, this
method is not restricted to semi-simple Lie groups only, but also
works for certain other groups, like some semi-direct product groups,
amongst which $G=\iso$, \cite{Rawnsley} \cite{Klaas}. 
We apply this method to construct the action for a free particle carrying a
Euclidean mass $r$ and a spin $s$, but we first recall the method in more
detail for a general Lie group $G$. 

A coadjoint orbit is obtained via the action of the group on a certain
coadjoint vector $Y\in \gog^*$, where $\gog^*$ denotes the dual of the
Lie algebra. It has a natural symplectic structure $\om$, and can be
quantised under Weil's integrality condition, which means that 
``$\om$ must be an integral element of the second cohomology group'', see
for instance \cite{Woodhouse}. So a coadjoint orbit can be regarded as the
phase space of some (classical) system, and geometrical quantisation leads
to a Hilbert space which forms a particular irreducible unitary
representation of the Lie group we started off with.  

To find the action for the classical system, we consider the canonical
one-form $\tht_{\mbox{\tiny{${Y}$}}}$ on the coadjoint orbit. Note
that if $\om$ corresponds to a nontrivial element of the second
cohomology group it cannot be exact, so the relation
$\om_{\mbox{\tiny{${Y}$}}}= -d\tht_{\mbox{\tiny{${Y}$}}}$ 
does not hold globally. This implies that a globally valid
expression for $\tht_Y$ cannot be obtained from the symplectic form
$\om_Y$ on the orbit, which follows directly from the Poisson--Lie
structure on $\gog^*$. 

As is well known the coadjoint orbit can be obtained as the reduced phase
space of a particle moving on the group \cite{AbrMars}. We denote
basis elements of $\gog$ by $T^a, a=1,...,n$, with
$n=\mbox{dim}(G)$, and the relations $[T^a, T^b] = {f^{ab}}_c T^c$, with
${f^{ab}}_c$ the structure constants. The dual basis on $\gog^*$ is
denoted by $\{t_a\}$, with $t_a(T^b) = {\dl_a}^b$. On the
full (unreduced) phase space $T^* G \simeq G \times \gog^*$ there is a
standard canonical one-form, defined via its contraction with a vector
field on $T^* G$
\be
\tht (x,Y) \cdot (X^{(x)}, Z) = Y(X^{(e)}) =: \langle \tilde{Y},
X^{(e)}\rangle 
\ee
Here $(X^{(x)}, Z)$ is an element of the tangent space to $T^* G$ in
the point $(x,Y)\in T^* G$, which is isomorphic to $T_x G \times
\gog^*$. The element $\tilde{Y}$ lying in $\gog$ is by definition dual
to $Y\in \gog^*$. For semi-simple groups it is simply the element
itself, because the inner product $\langle , \rangle$ is given by the
trace in the fundamental representation, $\langle T^a, T^b 
\rangle = \mbox{tr}(T^a T^b) = -\half \dl^{ab}$, identifying the algebra
with its dual. Note that for $G=\iso$, with the inner product on
the algebra as given in Eq.(\ref{eq:innerproduct}), this is not the case. 

The (left) action of $G$ on itself induces a symplectic action 
on $T^* G$, i.e.\ it leaves $\om$ invariant, and it
has the moment map $J\,:\, T^*G \to \gog^*$, given by
\be
\langle J\left( (x.Y , x)\right) , \xi\rangle = Y(\xi), \qquad Y
\in \gog^*, \xi \in \gog, x\in G,
\label{eq:momentmap}
\ee
with $x.Y$ denoting the coadjoint action of $x$ on $Y$. Moment maps are
directly related to the symmetries of the system, and it can be shown
\cite{AbrMars} that the reduced phase space, which is by construction a
symplectic manifold, is isomorphic to the coadjoint orbit 
\be
J^{-1}(Y) / N_{\mbox{\tiny{Y}}}\simeq G/N_{\mbox{\tiny{Y}}}\simeq
{\cal O}_{\mbox{\tiny{Y}}}, \quad Y\in \gog^*
\ee
where $N_{\mbox{\tiny{Y}}}= \{x \in G\,|\,x.Y = Y\}$ leaves the
coadjoint vector $Y$
invariant. Clearly, the particular choice of $Y \in {\cal
O}_{\mbox{\tiny{Y}}}\subset \gog^*$ is irrelevant. The orbit can therefore
be labeled by the set of invariants of $Y$ under the action of $G$, which
we will denote by $R$. We use the canonical one-form on $T^* G$ to
find a local expression for the one-form on the reduced phase space
$J^{-1}(Y) / N_{\mbox{\tiny{Y}}}$. For the moment map as given in
Eq.(\ref{eq:momentmap}) the pre-image of a certain coadjoint vector $Y$
is the trivial bundle over $G$, with in each element $x\in G$ the vector
$Y$ as the fiber. This bundle is isomorphic to the group $G$. The one-form on
the reduced phase space can be taken to be equal to the one-form on $T^*
G$, restricted to the pre-image of a certain fixed $Y \in \gog^*$, and
it is obviously invariant under the action of
$N_{\mbox{\tiny{Y}}}$. We conclude  that 
locally on the orbit ${\cal O}_R$ the canonical one-form is given by
\be
\tht_{\mbox{\tiny{Y}}}(x,Y) = \langle \tilde{Y},\xinv dx\rangle, \qquad
\mbox{fixed}\; Y \in \gog^*.
\label{eq:coadoneform}
\ee
 
In general, for a phase space with coordinates $\{\zeta\}$ (denoting
`positions' and `momenta') we may construct a first order Lagrangian
\cite{FadJac} 
\be
L\, dt = \tht(\zeta) - H(\zeta) dt,
\ee
with $H(\zeta)$ some Hamiltonian, and $\tht$ the canonical
one-form. For the simple example where 
the phase space is given by $T^* Q$, and $Q$ is $n$-dimensional, the
Lagrangian is given by $L = p_i \dot{q}^i - H(p,q), i=1,...,n$.

For the case of a coadjoint orbit we may take the canonical Hamiltonian
equal to zero, because at this point we are only interested in the
construction of the corresponding Hilbert space, and not in some possible
dynamics of the particle on the coadjoint orbit. The action is now given by
\be
S = \int L dt = \int \left(
\tht_{\mbox{\tiny{Y}}}(Y,x), \frac{d}{dt} \right) dt,
\label{eq:orbitaction}
\ee
with the large brackets $(,)$ denoting the contraction between the
one-forms and the vector fields on $\gog^*$. 

An element $\xi \in \gog$ is a linear function on the dual vector space,
and thus corresponds to a vector field $X_{\xi}$ on $\gog^*$. The 
symplectic two-form $\om_{\mbox{\tiny{Y}}}$ can be shown to be given by \cite{AbrMars}   
\be
\om_{\mbox{\tiny{Y}}} (X_{\xi}, X_{\eta}) = Y([\xi,\eta]).
\label{eq:coadjomega}
\ee
As expected, this corresponds precisely to the Poisson--Lie structure on
$\gog^*$:
\be
\{ F, G\}(Y) = Y ([dF_{\mbox{\tiny{Y}}}, dG_{\mbox{\tiny{Y}}}]).
\label{eq:poissonlie}
\ee
The differential $dF_{\mbox{\tiny{Y}}}$ of the function $F$ on $\gog^*$,
being a 
linear map from the tangent space to $\gog^*$ in the point $Y$, is
an element of $\gog^{**}\simeq \gog$, so $[ , ]$ is the Lie bracket, see
for instance \cite{Klaas}. It can be shown that the two-form
$\om_{\mbox{\tiny{Y}}}$ is
nondegerate on the coadjoint orbit, corresponding to the well-known
fact that the coadjoint orbits are the symplectic leaves of the dual of
the Lie algebra.  
 
Consider the coordinate functions on $\gog^*$
\be
Q^a = \langle Q, T^a\rangle,\quad Q= Q^a t_a \in \gog^*.
\label{eq:coordfunc}
\ee
They can be shown to generate the infinitesimal transformations $\dl Y =
T^a(Y)=Y(T^a)$, their Poisson brackets being given by 
\be
\{Q^a , Q^b\} = {f^{ab}}_c Q^c
\ee
In particular this holds for the coordinate functions on the orbit of
some fixed $Y$, where $Q = x.Y$, so they are the Noether charges
corresponding to the global symmetries on the coadjoint orbit. 

\subsection*{Coadjoint orbits of $\iso$}
For the case at hand, $G=\iso \simeq SU(2) \semdir \dubbelR^3 \simeq
SU(2)\semdir \gosu$. In computations we use the $4\times 4$ representation
of $\iso$, writing an element as
\be
U = \left(\begin{array}{cc} \Lm & \va \\ 0 & 1\end{array}\right)
\ee
with $\Lm \in SU(2)$ and $\va$ an element of the translation group. 
Denoting $U$ as $(\Lm, \va)$, the multiplication is given by   
\be
(\Lm_1, \vec{a}_1)(\Lm_2,\vec{a}_2) = (\Lm_1 \Lm_2, \va_1 + \Lm_1\va_2),
\label{eq:isomultiplication}
\ee
and the inverse by $(\Lm^{-1}, -\Lm^{-1}\va)$. The one-form $dU$ is
given by
\be
d(\Lm,\va) = \left(\begin{array}{cc} d\Lm & d\va \\ 0 & 
   0\end{array}\right).
\ee
The Lie algebra is $\goiso = \gosu \oplus \dubbelR^3$, its dual is
$\goiso^* = \gosu^* \oplus (\dubbelR^3)^* \simeq \dubbelR^3 \oplus
\dubbelR^3$. The coadjoint action of $\iso$ on $\goiso^*$ is denoted by
``dot'' and reads
\be
(\Lm, \va).(\vec{j},\vec{p}) = (\Lm \vec{j} +\va\wedge \Lm \vec{p}, 
\Lm \vec{p}), \qquad \vec{j}\in\gosu^*, \vec{p}\in \dubbelR^3,
\label{eq:coadaction}
\ee
where it is understood that $\Lm$ acts in the coadjoint representation on
$\vec{j}$. The invariants of the orbit of
$(\vec{j}, \vec{p})$ are the norm $|\vec{p}|$, and the inner product of
$\vec{p}$ and $\vec{j}$, both via the ordinary inner
product on $\dubbelR^3$.  We call these invariants $\vec{p}^2 = r^2,
\,r\geq 0$, and $\vec{p} \cdot \vec{j} = rs$. As a manifold the orbit
$(r,s), \,r>0$, is equivalent to the cotangent bundle $T^* S^2$, 
since the orbit of $\vec{p}$ is a 2-sphere, where in each point
$\vec{p}$ the $\vec{j}$ that satisfy $\vec{p}\cdot \vec{j} = rs$ span
an $\dubbelR^2$. For $r=0$ a coadjoint orbit is given by an
$SU(2)$ orbit of $\vec{j}$, so also an $S^2$. For a proof of
the fact that coadjoint orbits of semi-direct product groups indeed
have a nondegenerate symplectic structure we refer to \cite{Marsden} and
\cite{Klaas}. As can be concluded from the general derivation by
Rawnsley \cite{Rawnsley} the coadjoint orbits of $\iso$ 
can be quantised for all $r\geq 0$. Then $s\in \half\dubbelZ$ for
$r\neq 0$, and $s\in \half\dubbelZ$ with the restriction that the spin
$s\geq 0$ in case the Euclidean mass $r=0$, 
which indeed yields all irreducible representations of $\iso$. 

To obtain the irreducible
representation $(r,s)$ of $\iso$ we have to find a classical system
which has the coadjoint orbit ${\cal O}_{r,s}$ as its physical phase
space. The coadjoint element from which we construct the orbit is
$(\vj_0,\vep_0) = 
((s,0,0),(r,0,0))$, corresponding to $\tilde{Y}= r\cJ_0+s\cP_0$, due to the
form of the inner product on $\goiso$ as 
given in Eq.(\ref{eq:innerproduct}). From the general discussion above it
follows that the canonical one-form is given by
\be
\tht_{r,s}((\vj_0, \vep_0))= \langle r \cJ_0 + s\cP_0, U^{-1}dU \rangle. 
\ee
The second entry takes values in the Lie algebra of $\iso$, so it can
be written as 
\be
U^{-1} dU = \langle \Lm^{-1} d\va, \cP^b\rangle  \cP^b + \langle,
\Lm^{-1} d\Lm,  \cJ^b\rangle \cJ^b
\ee
where it is obvious in which algebra the inner products have to
be taken. Together with Eq.(\ref{eq:orbitaction}) this leads to the
free particle action
\be
S_p = \int dt \left( \vep \cdot \partial_t \va - \frac{s}{2}\mbox{tr}
\left( \Lm^{-1} \partial_t \Lm \cJ_0\right) \right). 
\label{eq:effaction}
\ee
The canonically conjugate momentum to $\va$ is defined by $\vep$
with components $p_b := r {\Lm_b}^0$, as will be explained shortly. 
We remark that the same action has been used by De Sousa Gerbert in
\cite{SousGer} for the case of (2+1)-dimensional gravity, where the
symmetry group is the Poincar\'e group in three dimensions,
$G=ISO(2,1)$.   

Let us take a closer look at Eq.(\ref{eq:effaction}). The first term
simply corresponds to a particle with position vector $\va$, momentum
$\vep$, and (Euclidean) mass $|\vep| = r$. The second term corresponds
to a pure spin $s$; it is nothing but the action of a system which has
the coadjoint orbit ${\cal O}_s$ of $SU(2)$ as its phase space (a
coadjoint orbit ${\cal O}_j$ of $SU(2)$ is a two-sphere of radius
$j$).  The
two terms are related, however, due to the fact that $\Lm$ is contained
in the definition of $\vep$. This is reflected in the Poisson brackets
of the symmetry generators, as can be understood as follows. 
From the general coadjoint orbit construction we know that the
symmetries of the action in Eq.(\ref{eq:effaction}) are generated by
the coordinate functions defined in Eq.(\ref{eq:coordfunc}). The
coadjoint element $Q = x.Y$ now corresponds to 
\be
(\vj, \vep) = U.(\vj_0,\vep_0) = (\Lm \vj_0 + \va \wedge\Lm\vep_0,
\Lm\vep_0),
\label{eq:freedoms}
\ee
leading to the coordinate functions
\bea
P_b &=& \langle\widetilde{(\vj_0,\vep_0)},\cP_b\rangle= r {\Lm_b}^0
\nn \\
J_b &=& \langle\widetilde{(\vj_0,\vep_0)},\cJ_b\rangle= s {\Lm_b}^0 +
{\ep_b}^{cd} a_c  (r {\Lm_d}^0) = (\va\wedge\vec{P})_b + \frac{s}{r} P_b.
\label{eq:pjfunct}
\eea
From the Poisson brackets of $\vec{P}$ with $\va$ with respect to the
variable $\va$ and its canonically conjugate momentum $\vep$, it
follows that $\vec{P}$ generates translations in $\va$, so it is simply 
equal to $\vep$. The second line of Eq.(\ref{eq:pjfunct}) can now be
identified as the angular momentum $\vj = \vec{l} + \vec{s}$, where
$\vec{l} = \va\wedge\vep$ is the orbital part, and $\vec{s}$ is the
intrinsic part. The latter is in fact the generator of the $SU(2)$
symmetry of the second term in the action. So we can say that
$\vec{P}= \vep$ and $\vec{J}=\vj$, with components $j^b =
\ep^{bcd} a_c p_d + \frac{s}{r} p^b$. By construction they satisfy the
Poisson brackets 
\be
\{p_a, p_b\} = 0,\quad \{j_a, p_b\}= \ep_{abc} p^c,\quad \{j_a, j_b\}
= \ep_{abc} j^c,
\label{eq:pbfuncs}
\ee
and the relations $p^a p_a = r^2$ and $p^a j_a = rs$. 

A derivation of the above Poisson relations via Dirac's theory of
constrained Hamiltonian systems is discussed in \cite{SousGer}.

For $r\neq 0$ the coadjoint orbit ${\cal O}_{r,s}$ is four
dimensional. One can 
either consider the $(\vj,\vep)$ with the constraints $|\vep|=r$ and
$\vep \cdot \vj = rs$ as the degrees of freedom, or the $U$ of
Eq.(\ref{eq:freedoms}), which are determined up to right
multiplication with elements $V$ in the two-dimensional subgroup of
$\iso$ that leave $(\vj_0,\vep_0)$
invariant. This two-dimensional centraliser consists of the abelian
semi-direct product group $U(1)\semdir T_t$ of rotations around
$\vep_0$ with translations in the `time' direction $T^0$. For $r=0$ the
coadjoint orbit ${\cal O}_{0,j}$ is the two-dimensional coadjoint
orbit ${\cal O}_j$ of $SU(2)$, corresponding to the $U\in \iso$ up to
right multiplication with elements of the four-dimensional centraliser
$SU(2)\semdir T$.

\subsection*{Coupling the particle to the gauge field}
Above we have derived the action for a free, spinning particle of
Euclidean mass $r$ and spin $s$. To introduce nonabelian interactions
for such particles they have to be coupled to the $\iso$ CS gauge
field. This is simply done via minimal substitution in the vector
field $\frac{d}{dt}$ in Eq.(\ref{eq:orbitaction}), using the component
of the gauge field in the direction of the Wilson line, which is the
world line of the particle
\be
\frac{d}{dt} \, \to\, D_t = \frac{d}{dt} + e^a_0 \cP_a + \om^a_0 \cJ_a.
\ee
The total action for the coupled particle thus becomes
\be
S_p + S_{\mbox{\scriptsize{int}}}= \int dt \left(p_b(\partial_t a^b+e^b_0
+ \ep^{bcd} \om_{c0} a_d) - \frac{s}{2}\mbox{tr}\left( \Lm^{-1}\left(
\partial_t + \om^b_0 \cJ_b\right)\Lm \cJ_0\right) \right),
\label{eq:totalaction}
\ee
from which we derive the following interaction term in the Lagrangian
\be
L_{\mbox{\scriptsize{int}}} = e^b_0 p_b + \om^b_0 j_b.
\ee
Adding the total action of Eq.(\ref{eq:totalaction}) to the gauge
field action of Eq.(\ref{eq:splitact}) now leads to the desired constraint
relations
\be
R_a(z) = p_a \,\dl^{(2)}(z - z_1),\quad  T_a(z)= j_a \,
\dl^{(2)}(z-z_1), 
\label{eq:nonvancurv}
\ee
instead of Eq.(\ref{eq:fis0}). This shows that indeed there is 
nonvanishing curvature at the position $z=z_1$ of the particle. In
particular, note that the Poisson brackets of the sources for the
curvature, as given in Eq.(\ref{eq:pbfuncs}), are consistent with the
Poisson brackets of the constraints as in Eq.(\ref{eq:RTbracks}).   

We now turn to the gauge symmetry of the model. It is well known that
CS theories are not invariant under gauge transformations at their
boundaries, and since punctures are part of the boundary of $\Sg$,
we may not allow for gauge transformations at their
positions. This means that effectively we only 
take gauge equivalence under the group of {\sl small} gauge
transformations, i.e.\ in Eq.(\ref{eq:phsptot}) we only divide out 
by these elements of the gauge group that take the value $e$ (unit in
$G$) at the punctures. Without this restriction the internal degrees of
freedom of a particle, corresponding to the $U$ (or $(\vj,\vep)$) of
Eq.(\ref{eq:freedoms}), could have been completely gauged away, reducing the
internal phase space to a single point, namely the unit element of $\iso$. 

\subsection*{The total phase space in the presence of a particle}
As our goal is to quantise the system of topologically interacting
particles, we should study the total phase space which includes both the
degrees of freedom of the gauge field, as well as the degrees of freedom of
the particles coupled to them. At first sight this seems to lead to quite an
elaborate analysis -- this phase space being the tensor product of ${\cal
M}_{\mbox{\scriptsize{phys}}}$ with the tensor product of the phase spaces
of each of the particles. The latter, in principle, for each
particle consists of the coordinates and momenta in space time,
together with the 
internal coadjoint orbit that leads to its nonabelian charges. However,
this total phase space can be largely simplified, as follows from the fact
that we have to impose the modified constraint equations from
Eq.(\ref{eq:nonvancurv}). We will now show that this basically allows one
to eliminate the internal degrees of freedom of the particles in favour of
the holonomy degrees of freedom of the gauge field.

By introducing the puncture we
have effectively created a noncontractable loop in $\Sg$, which contains
nonvanishing curvature as given by Eq.(\ref{eq:nonvancurv}). Before dividing
out the gauge symmetry as in Eq.(\ref{eq:phsptot}), or Eq.(\ref{eq:phsphom}),
the phase space of the gauge field in the presence of the particle 
consists of the holonomy   
\be
g_1:=w(C_1) = P \exp (\oint_{C_1} A_j dx^j) = \exp ( \int_{\sg} 
F_{ij} dx^i\,dx^j) = \exp (p^a \cJ_a + j^a \cP_a)\;\;\in\iso.
\label{eq:hol}
\ee
Here $C_1$ is a closed loop around the particle which encloses the surface
$\sg$. It follows that if we --for the moment-- ignore the gauge 
symmetry, i.e.\ we don't take conjugation equivalence under ${\cal G}(*)$
into account, we see that the classical state of the gauge field is given
by the $\iso$ group element $g_1 = w(C_1)$. Thus, the degrees of
freedom of the gauge field, as given by the holonomy in Eq.(\ref{eq:hol}),
are in direct correspondence with the internal degrees of freedom $(j^a,
p^a)$ of the particle, for a given Euclidean mass $r$ and spin $s$.

There is one subtlety, however, which concerns the following. It is
clear that due to the exponentiation in Eq.(\ref{eq:hol}) one
cannot distinguish between states of the gauge field that are related
to particles with Euclidean masses $r$, and states that are related to
particles with masses $r + 2\pi m$, with $m\in \dubbelN$. So upon
imposing the constraints the ``space of possible total phase spaces''
gets compactified in the $r$-direction, as opposed 
to the infinite range $r\geq 0$ for the coadjoint orbits, which correspond
to the internal phase spaces. In particular, states with Euclidean masses
$2\pi m$ cannot be distinguished from massless states. 

Concerning the spatial degrees of freedom we can say that for the case of a
single particle they completely 
decouple from the rest of phase space. In fact, the base point $*$ can be
chosen such that the particle is static at a certain point $z_1\in \Sg$. 

Let us now consider the question of the gauge symmetry. In fact, the
physical phase space of the 
gauge field as given in Eq.(\ref{eq:phsptot}) is invariant under (small)
gauge transformations. This means that we still have to divide the
classical single particle state given in Eq.(\ref{eq:hol}) by 
conjugation with ${\cal G}(*)$. Therefore, in the presence of just a single
particle the phase space ${\cal M}_1$ consists of only one point, with only
the labels of the conjugacy class of $g_1$ as nontrivial data. In other
words, all dynamical degrees of freedom are gauged away, leaving only the
gauge invariant quantities. For a general Lie group $G$ conjugation is in
fact the definition of the adjoint action in the Lie algebra, 
\be
\exp({\mbox{Ad}_g \xi}) := g (\exp \xi) \ginv, \quad g\in G, \xi \in
 \gog. 
\label{eq:adj}
\ee
This in turn determines the coadjoint action on the coordinates
\be
\exp(\mbox{Ad}_g (\xi^a T_a)) = \exp(\xi^a \mbox{Ad}_g T_a) = \exp(
 (\mbox{Co}_g \xi^a) T_a).
\label{eq:coadj}
\ee
At least for $SU(2)$ this makes clear that we can regard a conjugacy
class of the group as the exponentiation of a coadjoint orbit, where
again we see that the space of orbits becomes compactified. For $\iso$
it means that the coadjoint orbit with invariants $(r,s)$ corresponds to
the conjugacy class with invariants $(r$ mod $2\pi, s)$.

\subsection{The multi-particle case}

The situation becomes considerably more interesting if there is more than a
single particle present. As before, we impose the constraints from
Eq.(\ref{eq:nonvancurv}), effectively eliminating the internal degrees of
freedom of the particles in favour of the degrees of freedom of the gauge
field. The latter are purely contained in the choice of basis of the (based)
fundamental group $\pi_1(\Sg,*)$ of $\Sg$, and in the holonomies of the
flat gauge field around the basic loops.  

\begin{figure}[t]
\epsfxsize=7cm
\epsfysize=4cm
%\epsfangle=90
\centerline{\epsffile{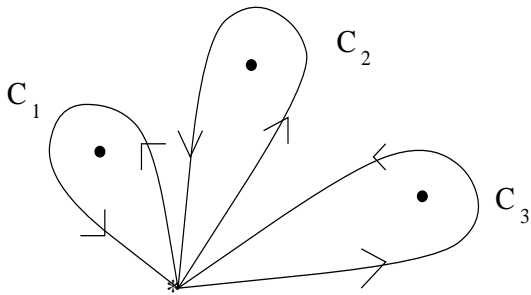}}
\caption{Choice of generators of $\pi_1(\Sg,*)$.}
\label{fig:loops}
\end{figure}

From the identification of the phase space ${\cal M}$ in 
Eq.(\ref{eq:phsphom}) we can see that it has a definite multiplicative
structure under the composition of $*$-based noncontractable loops in
$\Sg$. To give a well defined description of the phase space as given in
Eq.(\ref{eq:phsphom}) we must first define the generators of
$\pi_1(\Sg,*)$. Choosing another set of generators will give a different
characterisation of the phase space, but it doesn't change the physical
contents, of course.  We choose to order the punctures according to their
angle $\varphi_k$ (increasing in the clockwise direction)  
relative to the base point, and by their distance to the base point if they
have the same angle. The generators of $\pi_1(\Sg,*)$ are chosen to be the
homotopy classes of the noncontractable loops around single punctures,
which we denote by $C_1, C_2,..., C_N$, see figure \ref{fig:loops}. To each 
$C_i$ we assign a holonomy in the usual way, $g_i := w(C_i)$, as in
Eq.(\ref{eq:hol}), where by definition $C_i$ has to be traversed in the
counterclockwise direction. Traversing $C_i$ in the clockwise direction
yields the inverse holonomy. From now on we refer to the $C_i$ as the
`basic loops', instead of the `homotopy classes of the basic loops'.

Take $\Sg$ with $N$ punctures with nonvanishing curvature. Before  
dividing out the gauge symmetry the internal classical state will be
given by $(p^a_1, j^a_1,..., p^a_N, j^a_N)$. In the total phase space this
corresponds to holonomies $(g_1,..., g_N)$ with 
\be
g_k =  e^{p^a_k \cJ_a + j^a_k \cP_a}, \quad k=1,...,N.
\label{eq:gexp}
\ee
In the following we will in fact identify the individual holonomies
with the (classical states of the) individual particles. That is, we will
often refer to $g_k$ as `the $k$-th particle'. 

We will now discuss two characteristic features of the classical
multi-particle phase space, which we expect --at least partially-- to
remain valid on the quantum level. They are `fusion' and `braiding'.

\subsection*{Classical fusion} \label{ssclasfusion}

Firstly, consider the case where there are two particles present. 
In a given ordering one can simply determine the over-all holonomy,
because by traversing $C_{1\cdot 2} := C_1\circ C_2$ one 
determines the properties of the effective particle within the
loop, which consists of the two individual punctures (in the given
ordering). So the total holonomy is $g_{1\cdot 2} = w(C_{1\cdot 2})= g_1
g_2$, which  corresponds to {\sl fusion} of particle 1 and 2. This is the
definite --in principle nonabelian-- multiplicative structure of phase
space mentioned above.  

We call $g_3 :=g_{1\cdot 2}$. So the classical state of the total
(effective) particle, $g_3$, is uniquely determined by the classical states
of the particles associated with the basic loops. In particular, the
invariants $r_3$ and $s_3$ that label the conjugacy class of $g_3$ are
uniquely determined by the group multiplication. 

In case there are $N$ punctures, $\Sg$ is noncompact, and we take
the set $\{C_i\}_{i=1,...,N}$ defined above as the basis for
$\pi_1(\Sg,*)$, we will find the over-all holonomy 
\be
g_{\mbox{\scriptsize{tot}}}=\prod_{i=1}^N g_i.
\ee
Note that gauge transformations are not allowed at infinity
(they are not a symmetry of the CS-action), so we can shrink
infinity to a point and consider $(N+1)$ punctures on the
compactified $\Sg$, with the condition that $\prod_{i=1}^{N+1}
g_i=e$, where $e$ denotes the unit element in the group.

\subsection*{Classical braiding}

A second essential feature of topologically interacting particles consists
of the kinematical relations which their internal degrees of freedom have
to satisfy, if one considers the trajectories of the particles in space 
time. In two spatial dimensions these relations can be nontrivial, due to
the fact that the configuration space of distinguishable particles moving
in the plane is non-simply connected.  

\begin{figure}[t]
\epsfxsize=12cm
\epsfysize=4cm
%\epsfangle=90
\centerline{\epsffile{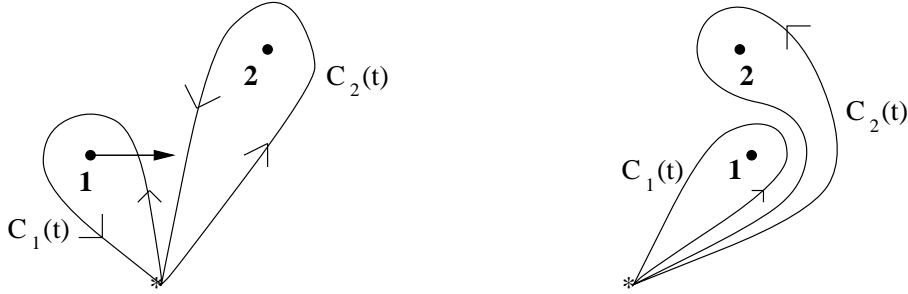}}
\caption{Basic loops at time $t$.}
\label{fig:braida}
\end{figure}

Consider the case of two particles. At a certain initial time $t$,
for a given ordering of the basic loops $C_1(t)$ and $C_2(t$), the classical
state is given by $(g_1, g_2)$, as depicted in figure \ref{fig:braida}. 
\begin{figure}[t]
\epsfxsize=11.5cm
\epsfysize=4.5cm
\centerline{\epsffile{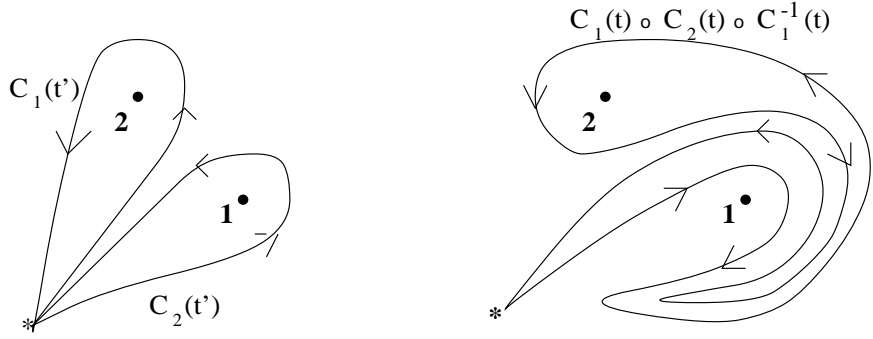}}
\caption{Basic loops at time $t'$ and their decomposition in loops at
time $t$.}
\label{fig:braidb}
\end{figure}
Suppose that we carry particle 1 in the $z_x$-direction (one of the
two independent 
directions on $\Sg$, which we parametrise by $z$), relatively to particle
2. At some final time $t' > t$ the configuration of the
particles will be as in Figure \ref{fig:braidb}. 
According to the chosen ordering of the basic loops  $C_1(t')$ will now
encircle particle 2, and loop $C_2(t')$ will go around particle 1. To
relate the holonomies of the final configuration to the holonomies of
the initial configuration we must express $C_1(t')$ and $C_2(t')$ in
$C_1(t)$ and $C_2(t$). It is clear that $C_2(t')=C_1(t)$. The loop
$C_1(t')$ can be decomposed in the loops $C_1(t)$ and $C_2(t)$  
\be
C_1(t') = C_1(t) \circ C_2(t) \circ C_1^{-1}(t),
\label{eq:classbraid}
\end{equation}
from which it follows that the corresponding holonomy is given by $g_1
g_2 \ginv_1$. Thus a counter clockwise interchange of the particles
can be described by
\be
{\cal R} (g_1, g_2) = (g_1 g_2 \ginv_1, g_1). 
\label{eq:br}
\end{equation}
If particle 1 had  
passed particle 2 above, instead of underneath, the interchange would have
been clockwise, and the first holonomy would have been conjugated by the
inverse of the second one, i.e.\ we would have ended up with the
fluxes $(g_2, \ginv_2 g_1 g_2)$. Note that in the initial as well as
the final state fusion leads to the same result, since $g_1 g_2 = (g_1
g_2 \ginv_1) g_1 = g_2 (\ginv_2 g_1 g_2)$, so the total holonomy
remains constant. This concludes
our description of braiding and fusion on the classical level.

\vspace*{.3cm}

To take the gauge symmetry into account we must divide out the
simultaneous conjugation of all holonomies. From now on we take $\Sg$
to be compact. Let us, for simplicity, consider again the case of three
punctures, 
where one of them represents spatial infinity. By a gauge transformation
we can put one of the holonomies in the simple form of a diagonal rotation
matrix and a pure time translation, we call this form `diagonal' as
well. However, the  
`relative differences' between holonomies are invariant, by which we mean
the following. Like on a coadjoint orbit, any element of a conjugacy class
can be written as a transformation of a representative element under the
group action. In Eq.(\ref{eq:freedoms}) an arbitrary element $(\vj, \vep)$
has been written as the transformation of the element $(\vj_0,\vep_0)$ by
the group element $U \in \iso$. For any $(\vj,\vep)$ and a fixed
$(\vj_0,\vep_0)$ the element $U$ is determined up to right multiplication by
an element $V\in \iso$ which leaves $(\vj_0,\vep_0)$ invariant. Similarly, 
any element $g$ in a conjugacy class $C_{r,s}$ can be written as the
conjugation of a representative element in that class. We choose this
representative element to be the diagonal element $g_{r,s} = \exp (r
\cJ_0 + s \cP_0)$, so we write $g = U g_{r,s} U^{-1}$. As explained in the
previous subsection for a single particle this $U$ can always be gauged
away, and it does not contain any dynamical information. However, in the
multi-particle case a gauge transformation acts simultaneously on all
holonomies, which must satisfy the `fusion constraint' we described above.
We write the fusion as 
\be
U_1 g_{r_1, s_1} U^{-1}_1\,\, U_2 g_{r_2, s_2} U^{-1}_2 := U_3 g_{r_3,
s_3} U^{-1}_3.
\label{eq:isofusion}
\ee
From this it is clear that simultaneous conjugation of all holonomies
will leave the relative differences $U^{-1}_1 U_2,\,\, U^{-1}_2 U_3$ and
$U^{-1}_3 U_1$ invariant, and that they can contain dynamical degrees of
freedom. 

Finally, to specify a certain classical state in the case of $N$
punctures on a noncompact space $\Sg$, we must define the ordering of the
punctures in space, and for each of them give the corresponding
holonomy. Per puncture the holonomy gives four degrees of freedom, since 
we consider the $r$ and $s$ to be fixed. Also, we must divide out the gauge
symmetry. This leaves us with a $4N-4$ dimensional phase space for $N\geq
1$, plus the choice of the ordering of the basic loops. As argued before,
for $N=1$ the phase space is a single point.    

\section{The quantum theory} \label{squant}
In the previous section we have determined the classical physical phase
space in the presence of particles. Now we turn to the problem of
quantisation.  
It is important to observe the natural cotangent bundle structure of
the phase space 
\bea
{\cal M}_{\mbox{\scriptsize{phys}}} &=& T^*{\cal N} 
\label{eq:multiphsp}\\
{\cal N} &=& \mbox{Hom}(\pi_1(\Sg,*); SU(2))/\sim
\label{eq:configspace}
\eea
where $\sim$ now denotes equivalence under conjugation by $SU(2)$ elements.
This makes a choice of polarisation (i.e.\ choice of `momenta' and 
`coordinates' inside phase space) straightforward. ${\cal N}$ is the
configuration space, where we emphasise that it has the unusual
property that its elements do not commute. Upon quantisation, the 
Hilbert space consists of square integrable functions on this
configuration space.

\subsection*{Parametrisations of $SU(2)$}
To discuss the quantisation in more detail we first fix some convenient
parametrisations of $SU(2)$. In the Euler--angle parametrisation each $g\in
SU(2)$ can be written as
\be
g = g_{\phi} a_{\tht} g_{\psi}
\label{eq:euler}
\ee
with
\be
g_{\phi} = \left(\begin{array}{cc} e^{\half i \phi} & 0 \\ 0 &
e^{-\half i \phi}\end{array}\right) , \qquad
a_{\tht} = \left(\begin{array}{cc} \cos \half\theta & -\sin \half\theta \\ 
\sin \half\theta & \cos \half\theta \end{array}\right) \nn
\label{eq:eulermatrices}
\ee
\be
0\leq\tht\leq \pi,\quad 0\leq\phi<2\pi,\quad -2\pi\leq\psi\leq 2\pi.\nn
\ee
Of course, all elements $g_{\phi}$ form the diagonal $U(1)$ subgroup.

Conjugacy classes of $SU(2)$ consist of all elements with the same angle 
of rotation $r$, in arbitrary directions, and thus can be thought of as
2-spheres with radius $r$. We take the diagonal element $g_r$
to be the representative of the conjugacy class 
\be
C_r := \{ x g_r\xinv\,|\, x\in SU(2)\},\quad 0\leq r\leq 2\pi.
\label{eq:conjr}
\ee
Let $0<r<2\pi$, then $C_r$ clearly consists of the elements
\be
g(r,\tht,\phi) := g_{\phi} a_{\tht} g_r a_{\tht}^{-1} g_{\phi}^{-1}. 
\label{eq:gagag}
\ee

We call the `centraliser' of an element the subgroup which leaves that
element invariant. Elements in the same conjugacy class have
isomorphic centralisers, so we can speak about {\sl the} centraliser
$N_r$ of the conjugacy class $C_r$, which we define 
as the centraliser of the representative of the conjugacy
class. This means that for $0<r<2\pi$ we have $N_r = U(1) = \{g_{\phi}\,|\,
-2\pi\leq \phi\leq 2\pi\}$, and its irreps are labeled by integers and
half-integers. For $r=0, 2\pi$ the  
centraliser is $SU(2)$ itself, with the irreps labeled by $j=0,\half,1,...$
Comparison of Eq.(\ref{eq:gagag}) with Eq.(\ref{eq:conjr}) shows that
$x$ actually denotes the 
direction within the conjugacy class $C_r$, so on the 2-sphere, and 
that it corresponds to the coset $g_{\phi}a_{\tht} N_r$.  
Take $\vec{\tau}$ to be the vector of generators of $SU(2)$ in the
fundamental representation, with $\tau_3$ diagonal. Then $g(r,\tht,\phi)$
can also be written as
\be
g(r,\tht,\phi) = \exp(i\frac{r}{2} \hat{n}(\tht,\phi)\cdot\vec{\tau})=
\unit \cos\frac{r}{2} + i\,\hat{n}\cdot\vec{\tau}\, \sin\frac{r}{2}
\label{eq:exponent}
\ee
so the representative of the aforementioned coset $xN_r$ is in 1--1
correspondence with the unit vector $\hat{n}(\tht,\phi)$, and we will
choose it to be $x=g_{\phi} a_{\tht}$. 
The unit and minus unit element correspond to $r=0$ and $r=2\pi$ 
respectively.

\subsection{Quantising the single particle system} \label{ssingleparticle}

In section \ref{s2a} we introduced particles, and showed that this leads
to the constraints of Eq.(\ref{eq:nonvancurv}). They imply that for the
case of a single particle the phase space as given in
Eq.(\ref{eq:multiphsp}) is the conjugacy class $C_{r,s}$, instead of the
coadjoint orbit ${\cal O}_{r,s}$ for the internal phase space of the
particle. In principle we should quantise the conjugacy class
$C_{r,s}$ directly, and the resulting physical states should reflect the
difference between $C_{r,s}$ and ${\cal O}_{r,s}$. However, we know the
mapping between the two spaces (the exponential mapping), and it is
clear that the group action is the same on each of them, as can be seen in 
Eqs.(\ref{eq:adj}) and (\ref{eq:coadj}). Therefore we will now
first quantise the coadjoint orbit, and simply accomodate for the
difference by making the appropriate identifications in the space of labels
of irreps. Then we identify the resulting Hilbert space as the
representation space of a `new' kind of underlying symmetry, the true
benefit of which will become fully clear if we study the multi-particle
case in the next section. 

The coadjoint orbit ${\cal O}_{r,s}$ can be geometrically
quantised for $r>0$ iff $s\in\half\dubbelZ$, and for $r=0$ iff
$s\in\half\dubbelZ$ with $s>0$. In particular, its Poisson structure is
given by 
\be
\{F,G\} (\vec{j}, \vec{p}) = \left\langle \vec{j}, \left[\frac{\partial
F}{\partial \vec{j}}, \frac{\partial G}{\partial\vec{j}} \right]\right
\rangle  + \left\langle \vec{p}, \rho'\left(\frac{\partial F}{\partial
\vec{j}}\right) \frac{\partial G}{\partial\vec{p}}\right\rangle -
\left\langle  \vec{p}, \rho'\left(\frac{\partial G}{\partial
\vec{j}}\right) \frac{\partial F}{\partial\vec{p}}\right\rangle
\label{eq:poisson}
\ee
see Marsden {\sl et al} in \cite{Marsden}, and also \cite{Klaas}. 
In Eq.(\ref{eq:poisson}) the functional derivative $\frac{\partial
F}{\partial \vec{j}}$ is regarded as an element of $\gosu$, and
$\frac{\partial F}{\partial\vec{p}} \in \dubbelR^3$. Also, $\langle\; ,
\;\rangle$ denotes both the pairing between $\gosu$ and $\gosu^*$, and
the inner product on $\dubbelR^3$. The $\rho'$ is the induced Lie algebra
representation. 
 
By construction quantisation of this orbit leads to the irreducible
unitary representation $(r,s)$ of $\iso$. The carrier space 
(Hilbert space) of the corresponding representation is given by 
\be
{\cal H}^{r,s} = L^2({\cal O}_r) \otimes {\cal H}_s,
\label{eq:carrier}
\ee
where ${\cal O}_r$ is the orbit of $\vep_0$ under the coadjoint action of
$SU(2)$, so the 2-sphere of radius $|\vep_0| = r$. The Hilbert space ${\cal
H}_s$ is the carrier space of the irreducible unitary representation $\Pi_s$
of the centraliser $N_r$. 

The action of the group in this representation is usually described as
follows. Let $\sg$ be a Borel section for $SU(2)/U(1) \simeq {\cal O}_r$, 
meaning a (Borel) mapping $\sg\,:\,{\cal O}_r \to SU(2)$ such that
$\sg(\vep).\vep_0 = \vep$ (again, `dot' denotes the (coadjoint)
action of $\sg(\vep) \in SU(2)$ on the element $\vep_0
\in \dubbelR^3$, we will omit it from now on). In other words, the section
$\sg$ assigns to a vector 
$\vep \in \dubbelR^3$ an element $x\in SU(2)$ which rotates the
representative element $\vep_0$ to $\vep$. If one chooses $\vep_0$ along
the $z$-axis, then $x$ corresponds to the direction $(\tht_p,\phi_p)$ of 
$\vep$. The action of an element $(\Lm,\va)$ on an element $\psi$ of the
space of Eq.(\ref{eq:carrier}) is given by  
\bea
\left((\Lm, \va) \psi^{r,s}_{\sg}\right)(\vep) &=& e^{i\vep\cdot\va}
\Pi_s(\sg(\vep)^{-1} \Lm \sg(\Lm^{-1}\vep))\,
\psi^{r,s}_{\sg}(\Lm^{-1}\vep) \nn\\
&=&  e^{i\vep\cdot\va}\Pi_s(\xinv \Lm x')\, \psi^{r,s}_{\sg}(\vep\,'),
\label{eq:action1}
\eea
where $\vep\,' = \Lm^{-1}\vep,\,\, x' = \sg(\vep\,')$.
For later applications we note that the space in
Eq.(\ref{eq:carrier}) is equivalent to the following space  
\be
L^2_s(SU(2), {\cal H}_s) := \left\{ \phi\,:\, SU(2) \to {\cal H}_s
\,|\,\phi(xh) = \Pi_s(\hinv) \phi(x), \forall h\in N_r \right\},
\label{eq:hilbtwo}
\ee
which equals the space of sections of an ${\cal H}_s$-bundle over
the orbit ${\cal O}_r$. In this case the action of $\iso$ simplifies to
\be
\left((\Lm, \va) \phi^{r,s}\right)(x) = e^{i\vep\cdot\va}
\phi^{r,s}(\Lm^{-1}x), \qquad \vep = x\,\vep_0.
\label{eq:alteraction}
\ee
We apologise for the slightly misleading notation, where $x$ denotes the
direction of $\vec{p}$, and has nothing to do with a possible canonically
conjugate variable to $\vep$. 

This concludes the quantisation of the internal space for the single
particle case. The Hilbert space is given in Eq.(\ref{eq:carrier}), or
equivalently in Eq.(\ref{eq:hilbtwo}). It has an $\iso$ symmetry, and
the action of the group is given in Eq.(\ref{eq:action1}), or
Eq.(\ref{eq:alteraction}). 

\vspace*{.3cm}

So far we have considered the quantisation of the coadjoint orbits of
$\iso$, whereas, as we pointed out before, we are actually interested in
quantising the conjugacy classes. Recall that the group action is the same
in both cases. We will now argue that quantisation of the conjugacy
classes leads to the same set of functions that form the Hilbert space of
Eq.(\ref{eq:hilbtwo}), where two important remarks have to be made. The
first one is that the label $r$ has 
to be restricted to the interval $0\leq r\leq 2\pi$, reflecting the
physical fact that the Euclidean masses are only defined modulo $2\pi$. 
An important result of this paper is that this compactification of $r$,
with the fact that the group action is still valid, {\sl does} have a
representation-theoretical foundation. This leads to the second remark,
which is that the functions of Eq.(\ref{eq:hilbtwo}) may 
be interpreted differently, namely as representations of a new algebraic
object, being the {\sl quantum double} of $SU(2)$, rather than as
representations of $\iso$. This quantum double of $SU(2)$ is denoted by
${\cal D}(SU(2))$, and we will show that the set    
\be
\{\mbox{conjugacy classes of}\,\, SU(2)\} \times 
\{\mbox{centraliser representations}\}
\ee
corresponds exactly to the complete set of irreducible unitary
representations of ${\cal D}(SU(2))$. Also, the $\iso$-action is covered by
the action of $\DS$, as will be explicitely shown in section \ref{s4}. 

For the single particle case this observation is not of crucial importance,
since the $\iso$ representation theory with compactification of the
$p^a$-space is sufficient to describe the physical Hilbert space. However,
for the multi-particle case this new symmetry is explicitly needed, because
the algebraic structures of the quantum double exactly cover the physical
properties of multi-particle systems with topological interactions, as will
be shown in the following sections. The fact that our physical Hilbert
space forms an irreducible unitary representation of the quantum double
of $SU(2)$ provides a direct answer to the problem of quantisation of a
conjugacy class in $\iso$, but a direct mathematical derivation will not
be given here
\footnote{Technically, it would be interesting to study the
connection between deformation quantisation of $iso(3)$ as a Lie
bi-algebra, and the quantum double $\DS$. In particular the role of
the deformation parameter is not clear in this case, since it is not
manifest in $\DS$.}. 

\vspace*{.3cm}

Let us briefly consider a possible characterisation of states in the 
Hilbert space we have found. 
From the description of the Hilbert space as in Eq.(\ref{eq:hilbtwo})
we see that for the single particle case a quantum state of the gauge
field can be written as a function $\phi^{r,s}$ on the group $SU(2)$, with
$N_r$-covariance under right multiplication of its argument. This is
effectively the same as a function on a conjugacy class $r$. In section
\ref{s4} we will give an explicit  
basis for the Hilbert space in terms of Wigner functions. At this point
it suffices to remark that one may choose an alternative basis of
`delta-sections' on the group, i.e.\ delta functions which by definition
have the covariance property 
\be
\dl_x^s (y) = \dl_e^s (\xinv y),\qquad \dl_x^s(yh) = \dl_{x\hinv}^s(y) =
\Pi_s(\hinv)\dl^s_x(y).
\label{eq:deltasections}
\ee
In bra-ket notation we can denote the (quantum) state corresponding to this
wave function by $|x\rangle_{r,s}$. This state is completely localised on
the homogeneous part of the group, and corresponds to the classical
holonomy   
\be
g = e^{p_a\cJ^a} = e^{\mbox{\scriptsize{Ad}}_x(r \cJ_0)} = x g_r
\xinv,\quad x\in SU(2). 
\label{eq:su2exp}
\ee
Alternatively, the state may be denoted by $|g\rangle_s$. In analogy with
the Aharonov--Bohm case we call $g\in SU(2)$ the  
{\sl flux} of the particle, and $s$ its {\sl charge}. 
As already remarked in the previous section, due to the gauge
equivalence in Eq.(\ref{eq:configspace}) the $x$ will not be a
dynamical degree of freedom in the single particle case. A gauge
transformation can always diagonalise a single holonomy, leaving only $r$
and $s$ as (gauge) invariants, which therefore constitute the physical
observables.   
\subsection{Quantising the interacting multi-particle system}
In section \ref{s2a} we pointed out that most of the interesting features of
the model only become manifest in the multi-particle context, and we are
faced with the problem of implementing them on the quantum level. The
conventional approach, where one uses the tensor product space of the free
particles as a starting point for a perturbative scheme, is not suitable for
the current situation, because topological interactions are nonperturbative
in nature. Also, a canonical analysis constructing the symplectic
structure of the multi-particle phase space of Eq.(\ref{eq:phsphom}) is
far from straightforward. We will now study how braiding and fusion of the
particles, which have nontrivial effects on the classical states, are
reflected on the quantum level, and how this in turn fixes the
appropriate (mathematical) framework. 

\subsection*{Quantum fusion}

Given two particles on $\Sg$, with fixed Euclidean masses $r_1, r_2\neq
0,2\pi$ and spins $s_1, s_2$. We take them each to be in the flux eigenstate
$|g_i\rangle_{s_i} = | x_i g_{r_i} \xinv_i \rangle_{s_i},\,\,i=1,2$, with
$x_i \in SU(2)$, and $g_{r_i}$ the representative, diagonal element in
conjugacy class $C_{r_i}$ as in Eq.(\ref{eq:eulermatrices}). So $g_i$
is the homogeneous part of the classical state of particle $i$. To the 
total holonomy, whose loop encloses both particles, we should assign a
quantum state, which must also correspond to the homogeneous part of its
classical counterpart. From the multiplication in $\iso$ we see that it is
given by $|g_3\rangle = |g_1 g_2\rangle$. On the quantum level
fusion can be written as
\be
|g_1\rangle\otimes|g_2\rangle =  |g_3\rangle.
\ee
What do we know of the state $|g_3\rangle$? Firstly, its Euclidean mass
$r_3$ depends on $r_1$ and $r_2$, and on the relative difference $\xinv_1
x_2$, as follows from the constraint
\be
x_1 g_{r_1} \xinv_1 x_2 g_{r_2} \xinv_2 = x_3 g_{r_3} \xinv_3. 
\label{eq:xxx}
\ee
Secondly, the $x_3$ corresponding to $g_3$ is determined by the group
multiplication in $SU(2)$. Recall that we can also write $g_k = \exp
(\vep_k\cdot \vec{\cJ}) = x_k g_{r_k} \xinv_k$. It follows that in
general, for noncommuting $g_1$ and $g_2$, this implies that $\vep_3$  
is {\sl not} the vector sum of $\vep_1$ and $\vep_2$. We return to
this explicit example in subsection \ref{ssversus}, but for the moment we
note the following. 

Despite the fact that states of individual particles correspond to
a subset of irreps of $\iso$, the two points above clearly show that
we are certainly {\sl not} dealing with ordinary $\iso$ tensor product
decomposition, as described for instance by Rno in \cite{Rno}, where he 
treats a.o.\ the Clebsch--Gordan coefficients for $E(3)= SO(3)\semdir
T(3)$. In the case of $E(3)$  a
state in an irrep $(r,s)$ is given by $|\vep\rangle$ with $|\vep|=r$, and
the tensor product state $|\vep_1\rangle \otimes |\vep_2\rangle \in (r_1,
s_1)\otimes (r_2,s_2)$ simply corresponds to the vector sum $|\vep_1 +
\vep_2\rangle$. Also, for $E(3)$ the resulting representation label $r_3$
lies in the interval $[|r_1-r_2|, r_1 + r_2]$, whereas in our case, due to
the fact that the tensor product state has to obey the group multiplication
rule, we have that $r_3$ is restricted to a smaller set.

\subsection*{Quantum braiding}

Counterclockwise interchange of the two particles must affect the quantum
state of the particles in a similar way as it affects the classical
state. This transformation is due to the fact
that for $n$ distinguishable particles moving in the plane the
configuration space $C_n = (\Sg^n - D)$, with $D$ the singular
configurations in which two or more particles coincide, is non-simply 
connected. This means that there is no unique way to quantise such 
multi-particle systems, but that it depends on the kinematics of the
particles, i.e.\ the braid formed by their world lines. More precisely,
there is a quantisation associated to each unitary irreducible
representation of the fundamental group of the configuration space,
$\pi_1(C_n)$, which for two spatial dimensions can be shown to be equal to
the pure braid group, which we denote by $B_n(\Sg)$. Thus the quantum
Hilbert space must (also) form an irreducible unitary representation of
$B_n(\Sg)$. This group is generated by 
the monodromy operators $\gm_{ij}$, which by definition transport particle
$i$ in a counterclockwise direction around particle $j$, without encircling
other particles. For more details we refer to \cite{WPB} and
references therein. 

The action of the monodromy operator $\gm_{12}$ is equal to twice performing
a counterclockwise interchange of two particles, which is denoted by ${\cal
R}^2$. A single counterclockwise interchange affects 
the classical state as described in Eq.(\ref{eq:br}), 
restricting it to the homogeneous parts we find that
\be
{\cal R} \,:\,|g_1\rangle |g_2\rangle \mapsto |g_1 g_2 \ginv_1\rangle
|g_1\rangle,\qquad g_i\in SU(2), i=1,2.
\label{eq:quantumbraiding}
\ee
This means that in general, for noncommuting $g_1$ and $g_2$, upon
interchange of the particles the wave function transforms in a nontrivial
way. This will certainly affect the results of scattering processes, and
gives rise to a nonabelian generalisation of the well known 
Aharonov--Bohm effect \cite{AhBo}. For the case of topological
interacting defects in broken gauge theories, where the residual
symmetry group is finite, this has been studied in detail in
\cite{WPB}. One of our objectives is to compute a two-particle scattering
amplitude for particles carrying charges of the continuous, nonabelian
group $\iso$, we come back to this in more detail in section \ref{s6}.

\vspace*{.3cm}

Summarising, we see that the states in our Hilbert space must satisfy
higly nontrivial combination rules, and transform under the action of the
braid group. These properties are to be interpreted as the manifestation
of a new type of
underlying symmetry which acts on the Hilbert space, and which we briefly
discussed towards the end of section \ref{ssingleparticle}. As was pointed
out there, for the case at hand this new symmetry corresponds to the 
quantum double of $SU(2)$, and we will study it in detail in the next
section.  
\section{The quantum double}\label{s3}
\setcounter{equation}{0}
A quantum double group --or simply `quantum double'-- is an example of a
quasitriangular Hopf algebra. We will briefly explain its most important
properties, and refer to the appendix for a summary of the basic
ingredients and properties.  

In general we denote a Hopf algebra by ${\cal A}$. It has a multiplication
$m$, which is a mapping   
\be
m\,:\,\cA \otimes \cA \to \cA.
\ee
It also has a comultiplication $\Dl$, which is a mapping 
\be
\Dl \,:\, \cA \to \cA \otimes \cA.
\ee
The fact that the quantum double is a {\sl quasitriangular} Hopf algebra
means that it contains an invertible element $R 
\in \cA\otimes \cA$ satisfying a set of relations, amongst which the
quantum Yang--Baxter equation on the space $V_1\otimes V_2 \otimes V_3$,
which reads
\be
R_{12} R_{13} R_{23} = R_{23} R_{13} R_{12}.
\ee
The $V_i$ are vector spaces on which ${\cal A}$ acts, and $R_{ij}$
denotes the representation of $R$ on the vectorspace $V_i\otimes V_j$, with
$R_{ij}:V_i\otimes V_j \to V_j\otimes V_i$.  

In the appendix we summarise the construction of the quantum double $\DG$
for a locally compact group $G$. Despite certain formal mathematical
objections \cite{KM}, we may stick to a straightforward generalisation of
the construction for finite $G$ and take
\be
\DG := C(G) \otimes \CC[G],
\label{eq:defdubbel}
\ee
where $G$ now denotes a (locally) compact Lie group. This means that we
consider $\DG$ as the tensor product of the algebra $C(G)$ of 
continuous functions on $G$, with the group algebra $\CC[G]$. A general
element of $\DG$ is denoted by $\fog$, with $f$ a (continous) function on
$G$, and $g$ in the group algebra $\CC[G]$.  
For our applications we don't run into difficulties with ill-defined
expressions, products of Dirac delta functions, etc, which is the
reason we can use the above definition for $\DG$, with the advantage that 
it makes a clear distinction between the actions of $C(G)$ and $\CC[G]$
within the quantum double action.
\subsection{The irreducible unitary representations of $\DG$}
The derivation and classification of the unitary irreducible representations
(irreps) of $\DG$ for locally compact $G$ has been given in \cite{KM}
in the mathematically precise formulation with $\DG = C(G\times
G)$. We recall a main result in the formulation with $\DG = C(G)
\otimes \CC[G]$. 

Let us first define the following space of functions
\bea
L_\alpha^2(G,V_\alpha)  := \{ \phi : G\to V_{\al} \;|\; \phi(xh) = 
\al (h^{-1})
\phi(x),\;\;\forall h\in N,\nn\\ 
\mbox{and} \quad
\|\phi\|^2:= \int_G \|\phi(x)\|_{V_\alpha}^2\,dx <\infty\}
\label{eq:hilbertspace}
\eea
Here, $N$ is a subgroup of $G$ acting on a Hilbert space $V_{\al}$
via an irreducible unitary representation $\al \in \hat{N}$, and $dx$
is the normalised Haar measure on $G$. 
By $\Conj(G)$ we denote the collection of conjugacy classes of the compact
group $G$. For each conjugacy class $C_A$ we now choose some
(representative) element $g_A\in C_A$, with $N_A$ its centraliser in $G$.
Also, we choose a unitary irrep $\alpha\in\widehat{N_A}$ of $N_A$ on some
Hilbert space $V_\alpha$.
The following theorem from \cite{KM} classifies the irreps of
$\DG$.
\begin{thm} 
For $A\in\Conj(G)$ and $\alpha\in\widehat{N_A}$ 
an irreducible unitary representation
$(A,\al)$ of $\DG$ on $L_\alpha^2(G,V_\alpha)$ is given by
\be
((\fog)\aphi_{\al})(x):= f(xg_A x^{-1})\,\aphi_{\al}(g^{-1}x),
\label{eq:repcomp}
\ee
and every  irrep of $\DG$ is equivalent to some $(A,\alpha)$. 
\label{thm:reps}
\end{thm}
In Eq.(\ref{eq:repcomp}) it is understood that
$\fog$ acts in the representation $(A,\al)$. We have used the more
physical convention of labeling the element of the representation
space, $\aphi_{\al}$, instead of the element of the quantum double, $\fog$. 

What is the physical content of Eq.(\ref{eq:repcomp})? We will soon work
it out in detail for our case of $\iso$ CS theory, but for now we
remark the following about the function $\aphi_{\al}$, and the action of
the quantum double element $\fog$. Consider a particle inserted at a
puncture in $\Sg$, which 
carries quantum numbers $(A,\al)$. These quantum numbers correspond to the
invariant $A$ of the holonomy $g$ around the puncture (its conjugacy class),
and the irrep $\al$ of the group that leaves $g$ invariant. The element 
$x\in G$, which is related to the holonomy via $g=x g_A \xinv$,
contains all internal configuration space degrees of freedom for the
gauge field in the presence of the particle at the puncture. The
$\aphi_{\al}(x)$ corresponds to the wavefunction of the total,
interacting system. We may describe the functions of
Eq.(\ref{eq:hilbertspace}) in the --more physical-- bra-ket notation, 
\be
\aphi_{\al, a}(x) := \langle x\,|\,(A,\al)\,a\rangle,
\label{eq:braket}
\ee
where $a$ denotes an element of the set of labels of the basis functions on
the conjugacy class. In Eq.(\ref{eq:ketsu2}) we give an explicit example for
the case of irreps of $\DS$. This notation will turn be useful in the tensor
product decomposition, but to describe the action of 
the quantum double the functions $\aphi_{\al}(x)$ are more suitable than the
kets $|(A,\al)\,a\rangle$. 

Concerning the element $\fog \in \DG$, we can distinguish two types of
actions in Eq.(\ref{eq:repcomp}). Firstly, the group-part of the quantum
double, i.e.\ $g$ in $\fog$, acts as an ordinary representation of $G$ on
the linear space $L^2(G)$ of square integrable functions on $G$. This is
extended to an action of the group algebra in the obvious way. Secondly,
the function-part of the quantum double, so $f$ in $\fog$, picks up the
holonomy $g = x g_A \xinv$. It is precisely this last
feature which will enable us to give a first quantised description of the
nontrivial properties of multi-particle systems, like the fusion and
braiding we have discussed in the previous sections. 

\section{The quantum double of $SU(2)$} \label{s4}
\setcounter{equation}{0}
In this section we explicitely construct the unitary irreducible
representations (irreps) of $\DS$, and point out that their representation
spaces correspond precisely to the total quantum Hilbert spaces that we
described in section \ref{ssingleparticle}. 

From the parametrisation of $SU(2)$ as given in Eq.(\ref{eq:gagag}) and
the classification of the irreps of $\DG$ for general compact $G$
we see that for ${\cal D}(SU(2))$ we can distinguish two classes of
irreps, as described in \cite{KM}. They differ by the subset of
$\Conj(SU(2))$ where $r$ belongs to, and correspond to two types of
particles: 
\vspace*{.2cm}\\
(i)  generic representations\\
For $0 < r < 2\pi$ the centraliser $N_r$ equals $U(1)$, and its irreps
are labeled by $n\in\half\dubbelZ$. Their representation spaces are
equal to $V_n = \CC$.  (We denote these irreps by $n\in \half
\dubbelZ$ to distinguish them from the irreps $(r,s)$ of $\iso$.)
The Hilbert space of the irreps $(r,n)$ of $\DS$ is given by
\be
V^r_n := L^2_n(SU(2)) = \{\phi:SU(2)\to \CC \;|\;\; \phi(x h) = 
n(\hinv) \phi(x), \quad \forall h\in U(1)\}.
\label{eq:suhilbsp}
\ee
Note that $n(h)$ here means `the element $h\in U(1)$ in the $n$-th
irreducible unitary representation of $U(1)$'. 

Physically these representations corresponds to particles with nonzero
Euclidean mass $r$ (up to $2\pi$), and spin $n$, which can take any integer
or half integer value. We emphasise that the variable $x$ denotes the
direction of the `Euclidean momentum' $\vep$, not a Euclidean space time
coordinate. 

\vspace*{.2cm}

We  now show that $\iso$ in a sense is contained in $\DS$, by
identifying which 
particular elements of $\DS$ correspond to particular elements of
$\iso$, in a single irreducible unitary representation. Consider a state
$\rphi_n$ in a fixed irrep $(r,n)$ of $\DS$. The 
action of $(\fog)\in \DS$ on this state reads 
\be
((\fog) \rphi_n) (x) = f(xg_r\xinv)\, \rphi(\ginv x).
\ee
Now take for $f$ the particular function 
\be
f_{\vec{a}}(\tilde{g}) := e^{i \vec{p}\cdot \vec{a}},\quad
\tilde{g}=xg_r\xinv \in SU(2), \,\, \vep = x \vep_0, \,\,|\vep_0| = r,
\ee
then the transformation of $\rphi_n$ is given by
\be
((f_{\vec{a}} \otimes g) \rphi_n)(x) = e^{i \vep\cdot \vec{a}}
\,\,\,\rphi_n(\ginv x). 
\ee
Compare this to Eq.(\ref{eq:alteraction}), which describes the
transformation of the same function $\rphi_n$ under the action of $\iso$ in
the irrep $(r,n)$, then it follows that 
\be
f_{\vec{a}}\otimes g \,\in \DS \,\Leftrightarrow \, (g,\vec{a})\,\in
\iso,\qquad g\in SU(2), \vec{a} \in \dubbelR^3.
\label{eq:isoindsu}
\ee
Thus we see that under a single irrep, i.e.\ on the one-particle level,
$\iso$ is contained in $\DS$.  
\vspace*{.15cm}\\
(ii)  special representations\\
When $r=0,2\pi$ the corresponding conjugacy class consists of only one 
element (namely the unit and minus the unit, respectively), and the
centraliser is $SU(2)$. By Remark 3.12 in \cite{KM} the corresponding
irrep of ${\cal D}(SU(2))$ can act directly
on $V^0_l=V^{2\pi}_l = \CC^{2l+1}$. However, the definition of the
representation spaces in general, Eq.(\ref{eq:hilbertspace}), tells us that
for the special representations we should consider the space $L^2_l(SU(2),
\CC^{2l+1})$. There is the following relation:
\bea
\CC^{2l+1} &\Longleftrightarrow & L^2_l(SU(2),\CC^{2l+1}) 
\label{eq:spechilb}\\
v &\Longleftrightarrow & \phi:x\mapsto \Gm_l (\xinv) v
\label{eq:hilbspaces}
\eea
and thus $v = \phi(e)$, and we should also label such a $\phi$ with the 
basis vector $v$ it is related to.  To avoid confusion with functions on 
the group we denote the $l$-th representation of $SU(2)$ by $\Gm_l$. 
Eq.(\ref{eq:hilbspaces}) follows from the 
defining property of the elements of the representation space as given in
Eq.(\ref{eq:hilbertspace}):
\bea
\phi(x\ginv) &=& \Gm_l(g) \phi(x),\qquad\qquad x,g \in SU(2) \nn\\
\mbox{take}\,\,g=x \Rightarrow \phi(e) &=& \Gm_l(x) \phi(x)
\eea
The special representations correspond to sources that have zero
Euclidean mass (again up to $2\pi$). They carry $SU(2)$ angular momentum,
in contrast to the $U(1)$ angular momentum of the generic representations.

\vspace*{.3cm}

There is a very convenient choice of basis of the Hilbert spaces for
all irreps of ${\cal D}(SU(2))$, corresponding to the {\sl Wigner
functions} $D^j_{mn}$, with restrictions on the labels for  
various distinct representations. Indeed, the Wigner functions span the 
algebra of functions on $SU(2)$, and later on their properties will allow
us to perform explicit calculations for the decomposition of tensor product 
representations of ${\cal D}(SU(2))$. Wigner functions
are well studied, we follow the notation of \cite{VMK}, and have made
extensive use of \cite{Vil}. For $x\in SU(2)$ parametrised by the Euler
angles as in Eq.(\ref{eq:euler})
the Wigner function $D^j_{mn}$ corresponding to the $m,n$-th matrix element
in the $j$-th irreducible representation takes the value
\be
D^j_{mn}(x) = e^{-im\phi} P^j_{mn}(\cos \tht) e^{-in\psi},
\label{eq:djmn}
\ee
where $P^j_{mn}$ are a subset of the Jacobi polynomials. For more details
we refer to section 3.3 in \cite{Vil}.
\vspace*{.2cm}\\
(i) Let us first consider again the generic representations. Recall that
any $g\in 
SU(2)$ in a given conjugacy class $g_r$ can be written as $g = xg_r\xinv = 
g_{\phi} a_{\tht} g_r \ginv_{\phi} a^{-1}_{\tht}$. 
The representation space from Eq.(\ref{eq:suhilbsp}) will be denoted by 
$V^r_n$, and the covariance property reads
\be
\phi(xh) = e^{-in\zeta} \phi(x), \quad \forall h=e^{i\zeta}\in U(1).
\ee
For clarity we denote the element of the centraliser by $h$ instead of
$e^{i\zeta}$. The conjugacy class representative will still be denoted by
$g_r$, although they are in the same centraliser $U(1)$.
From Eq.(\ref{eq:djmn}) we see that
\be
D^j_{mn} (xh) = e^{-in\zeta} D^j_{mn}(x),\qquad \forall h=e^{i\zeta}\in 
U(1).
\ee
This shows that the set $\{D^j_{mn}\,|\,n\,\,\mbox{fixed},\;j\geq n, 
-j\leq m\leq j\}$ has exactly the right covariance property, in other
words, they are indeed sections of $V_n$-bundles over $C_r\simeq
SU(2)/U(1)$.  

As the Wigner functions form a complete set on $SU(2)$, the aforementioned
set forms a basis for a Hilbert space corresponding to an irrep
of ${\cal D}(SU(2))$ with centraliser representation $n$
and an arbitrary conjugacy class $0<r<2\pi$. In other words, the Hilbert
spaces for irreps with the same $n$ and different $r$
are equivalent, and thus can be spanned by the same basis. To specify which
${\cal D}(SU(2))$-representation we are dealing with, the set 
$\{D^j_{mn}\,|\,n\,\,\mbox{fixed}\}$ should also be
given a label $r$ which denotes the behaviour under the action of 
${\cal D}(SU(2))$. We write $\rD^j_{mn}\in V^r_n$ with
\be
\left((\fog)\rD^j_{mn}\right) (x) = f(x g_r \xinv)\, \rD^j_{mn} (\ginv x).
\label{eq:qdaction}
\ee
An arbitrary element of the representation space can be written as
\be
\rphi_n (x)=\sum_{j\geq n}\sum_{-j\leq m\leq j}
c_{jm}\,\rD^j_{mn}(x),\quad x\in G.
\label{eq:su2phidecomp}
\ee
Note that the sum over $j$ is infinite, so the generic irreducible
unitary representations of $\DS$ are infinite
dimensional. In more physical terms we say
that the wavefunction for the system in the presence of a particle with
nonzero (up to $2\pi$) Euclidean mass and spin $n$ can be
expanded on the set of Wigner functions with a fixed label $n$ and $j\geq
n$. 

The bra-ket notation of Eq.(\ref{eq:braket}) for this case reads
\be
\rD^j_{mn}(x) = \langle x\,|\,(r,n)\,j m\rangle,
\label{eq:ketsu2}
\ee
so the label $a$ of Eq.(\ref{eq:braket}) here corresponds to the labels
$j,m$ of the basis functions on any conjugacy class $r\neq 0, 2\pi$. These
are nothing but the labels of the spherical harmonics, since the 
generic conjugacy classes are 2-spheres.
\vspace*{.15cm}\\
(ii) Next we consider the special representations. The representation space is
given in Eq.(\ref{eq:spechilb}). Let $v_n = v^{(l)}_n$ be the $n$-th basis 
vector in $\CC^{2l+1}$. Then the $m$-th component of the $(2l+1)$-dimensional 
basis function (or section of a fibre bundle over the group)
$\ophi_{l,n}$ is given by
\be
\left(\ophi_{l,n}\right)_m (x) = D^l_{mn}(\xinv).
\label{eq:specwign}
\ee
For clarity we will omit the labels $(0,l)$ when it is clear which
representation we are working in. 
%Note that a specific {\sl state} of the corresponding excitation is given 
%by the function in a specific point $x$. The (special) case of the pure
%state $v_n$ (which can be suggestively denoted by $|l,n\rangle$) is given by 
%$\left(\phi_n\right)_m (e) = \sum_m\dl_{mn} v_m = v_n$. We will come back 
%to this later when we consider the decomposition of tensor product states.
The label $n$ in fact has to be fixed, as a matter of choice of basis 
in the fibres over the group. This makes it more obvious that we are dealing
with ordinary $SU(2)$ representations, where $l$ denotes the irrep and
$m$ the state.
The action of ${\cal D}(SU(2))$ on this basis reads
\be
(\fog) \left(\phi_n\right)_m (x) = \left( (\fog) D^l_{mn}\right) (\xinv) = 
f(e)\, D^l_{mn} ((\ginv x)^{-1}) = f(e)\, D^l_{mn}(\xinv g).
\ee
The action of $SU(2) \subset {\cal D}(SU(2))$ indeed is the ordinary
group action on the $l$-th irreducible unitary representation, as can
be seen from the following arguments:
\bea
\psi_n(x) &:=& (1\otimes g) \phi_n (x) = \phi_n(\ginv x), \qquad
g\,\,\mbox{fixed}\nn \\
 &=& \Gm_l(\xinv) \Gm_l(g) v_n \nn\\
\psi_n &\Longleftrightarrow& \Gm_l (g) v_n.
\eea
This means that after the action of $(1\otimes g)$ on $\phi_n$ the
resulting function $\psi_n$ has to be expanded on the basis $\Gm_l (g)
v$, instead of the basis $v$. In terms of Wigner functions 
this is explicitly given by
\be 
\sum_m (1\otimes g)_{pm} D^l_{mn} (\xinv) = 
 \sum_{m'} D^l_{p m'}(\xinv) D^l_{m' n}(g),
\ee
and it shows that it is in fact a right action. The labels $(p,m)$ denote the
matrix element in the $(0,l)$-th irreducible representation of $\DS$. 
This description of the Hilbert spaces of the special irreps may at first
sight appear to be unnecessarily complicated, especially in view of the
equivalence in Eqs.(\ref{eq:spechilb})--(\ref{eq:hilbspaces}). However,
in calculating the decomposition of tensor products of irreps it will 
turn out to be convenient.

Finally, we remark that this Hilbert space can also be considered to be
spanned by the kets $\{ |(0,l)\,m\rangle \}$, $-l\leq m\leq l$, reflecting
the fact that the conjugacy class is just the unit, or minus unit
element, and the Hilbert space forms an irrep of the centraliser $SU(2)$. 
\section{Tensor product representations}\label{s5}
\setcounter{equation}{0}
In this section we will explain that the quantum double has precisely the
correct (Hopf) algebraic structures to describe topologically
interacting multi-particle systems. We start with a general remark on
tensor product actions. 

Suppose that we have two particles, each corresponding to an irrep of some 
underlying symmetry group $G$. To describe the transformation properties of
the combined system we must give the action of $G$ on the tensor product
space of the two representations. This is defined via the so-called
comultiplication $\Dl\,:\, G\otimes G$, which for ordinary groups is given
by $\Dl(g) = g \otimes g$, for any $g\in G$. For more complicated algebraic
structures which may arise as a symmetry of physical systems, like
the quantum double $\DG$, the comultiplication can be much less trivial.
For example, for the algebra of functions on a group $C(G)$ the
comultiplication is given in the Appendix in Eq.(\ref{eq:comfuncalg}). 
In general, it is the comultiplication which determines the decomposition
of tensor products into irreducible components, and we now turn to the case
of $\DG$.  

In the appendix the explicit form of the comultiplication of $\DG$ is given
in Eq.(\ref{eq:comdub}). Formally this
action has been derived in \cite{KBM}, we will rephrase it here in the
(physically) more convenient description with $\DG = C(G) \otimes \CC[G]$.
 
Let $(A,\al)$ and $(B,\bt)$ be irreducible unitary representations, then 
for the representation space of the tensor product
we take the following space of vector valued functions on $G\times G$ (for
simplicity we here omit the argument of the $L^2$-space)
\bea
L^2_{\al\bt} &:=& \{ \Phi : G\times G
\to V_{\al}\otimes V_{\bt}\;|\; \Phi(x_1 h_1, x_2 h_2) = \al (\hinv_1)
\otimes \bt (\hinv_2)\, \Phi (x_1, x_2), 
\label{eq:tensprodspace}
\\
& & \qquad \forall h_1\in N_A, h_2 \in N_B\,\,\} \nn 
\eea
and it is understood that the functions are square integrable. 
The action of $\DG$ in the representation $(A,\al)\otimes (B,\bt)$ is given
by  
\be
\left(\Delta(\fog)\,\ABPhi_{\al,\bt}\right)(x_1, x_2)
= f(x_1 g_A \xinv_1 x_2 g_B \xinv_2)\, \ABPhi_{\al,\bt}(\ginv
x_1,\ginv x_2), 
\label{eq:tensproduct}
\ee
where we have used the comultiplication of $\fog$ as given in 
Eq.(\ref{eq:comdub}). It is understood that $\fog$ acts in the representation
$(A,\al)\otimes (B,\bt)$ via the explicit labeling of the function
$\ABPhi_{\al,\bt}$.  The functions of the form
\be
\ABPhi_{\al,\bt}(x_1, x_2) = \aphi_{\al}(x_1)\otimes \bphi_{\bt}(x_2) 
\in V_{\al} \otimes V_{\bt}
\ee
with $\aphi_{\al}(x_1)$ and $\bphi_{\bt}(x_2)$  basis functions of 
the representation spaces for $(A,\al)$ and $(B,\bt)$ respectively, span a 
dense subspace of $L^2_{\al\bt}(G\times G, V_{\al}\otimes V_{\bt})$,
and the  positive--definite inner product reads
\be
\langle \Phi_1, \Phi_2\rangle := \int_G \int_G 
\langle{\aphi_1}_{\al}(x_1),{\aphi_2}_{\al}(x_1)\rangle_{V_{\al}}\,
\langle{\bphi_1}_{\bt}(x_2),{\bphi_2}_{\bt}(x_2)\rangle_{V_{\bt}}\,
dx_2\,dx_1.
\label{eq:tensinprod}
\ee
We have omitted the labels $(A,\al, B,\bt)$ on the functions $\Phi$, and 
will always do so when it is obvious which representation we are
working in.

\subsection{$\iso$ versus $\DS$} \label{ssversus}
In Eq.(\ref{eq:isoindsu}) we showed that the inhomogeneous group $\iso$ is
contained in the quantum double $\DS$, in the sense that within a single
irrep a subset of elements of $\DS$ can be identified with $\iso$. 
In this section, however, we will show explicitly that there exists a
major difference between $\iso$ and $\DS$, which is their action on
tensor product representations.  
 
To that aim, first consider the representation $(r_1,s_1)
\otimes (r_2,s_2)$ of $\iso$, with $r_1, r_2\neq~0$. The action of an
element $(g,\vec{a})$ is defined via its comultiplication
\be
\Dl(g,\vec{a}) = (g,\vec{a}) \otimes (g,\vec{a})
\ee
and it reads 
\be
\left(\Dl(g,\vec{a})\Phi\right)(x_1, x_2) = e^{i(\vep_1+\vep_2)\cdot\vec{a}}\,
\Phi(\ginv x_1, \ginv x_2),
\label{eq:isotensoraction}
\ee
where it is understood that $\Phi\in {\cal H}^{r_1,s_1}\otimes {\cal
H}^{r_2,s_2}$. The vectors $\vep_1$ and $\vep_2$ are defined via the action
of $x_1$ and $x_2$ respectively, as in Eq.(\ref{eq:alteraction}). 

Next, consider the representation $(r_1,n_1) \otimes
(r_2,n_2)$ of $\DS$, for $r_1, r_2\neq 0, 2\pi$, and the quantum double
element $f_{\vec{a}}\otimes g$, which is related to $(g,\vec{a})\in \iso$. 
With the tensor product action for the quantum double as given in
Eq.(\ref{eq:tensproduct}) we find that
\bea
\left(\Dl(f_{\vec{a}}\otimes g) \Phi\right)(x_1,x_2) &=& f_{\vec{a}}(x_1
g_{r_1} \xinv_1 x_2 g_{r_2} \xinv_2)\, \Phi(\ginv x_1, \ginv x_2)  \nn\\
&=& e^{i \vep_3\cdot \vec{a}}\,\Phi(\ginv x_1, \ginv x_2),
\label{eq:dsutensoraction}
\eea
with $x_3$ defined by the constraint given in Eq.(\ref{eq:xxx}), and the
vector $\vep_3 = x_3 (\vep_3)_0$, with $|\vep_3|=r_3$. 

Comparison of Eqs.(\ref{eq:isotensoraction}) and (\ref{eq:dsutensoraction}) 
explicitely shows the difference between $\iso$ and $\DS$ under tensor
product representations. It also makes clear that the quantum double
takes the fusion process described in section \ref{squant} properly into
account, whereas the inhomogeneous group does not. This becomes even 
more obvious if one studies the decomposition of tensor products into
single irreps, which we now turn to.

\subsection{Fusion rules and Clebsch--Gordan coefficients}
In general the decomposition of the tensor products of two irreps into
single irreps is described by the Clebsch--Gordan series, which can be
denoted by 
\be
(A,\al) \otimes (B,\bt) \simeq \bigoplus_{\gm} \int^{\oplus} \Nabc
\,\,(C,\gm)\,\,d\mu(C).
\label{eq:fusion}
\ee 
Here $\mu$ denotes in fact an equivalence class of measures on the set of
conjugacy classes, we refer to \cite{KBM} for more explanation. The
coefficients $\Nabc$ are called the `multiplicities', or `fusion rules'.  

When dealing with ordinary groups one may calculate the multiplicities via
the characters of the irreps
\be
n^c_{ab} = \int_{h\in H} \overline{\chi^c(h)} \chi^a(h) \chi^b(h)\,dh.
\ee
However, for technical reasons this cannot be generalised straightforwardly
for the (infinite dimensional) irreps of the quantum double.
To obtain an expression for the multiplicities $\Nabc$ 
we construct the projection of a tensor product state onto the
direct integral/sum of Hilbert spaces in the decomposition of
Eq.(\ref{eq:fusion}). We subsequently compare the squared norms of the
ingoing state with the direct integral/sum of squared norms in the
irreducible Hilbert spaces, and thus deduce an implicit expression for the
multiplicities. The construction of this projection has been derived in
\cite{KBM}, here we recall the result.

\vspace*{.2cm}

We fix the irreducible unitary representations $(A,\al), (B,\bt)$, and
$(C,\gm)$, and take $\xi \in N_A\backslash G/N_B$. The representation
spaces $V_{\al}, V_{\bt}$ and $V_{\gm}$ have dimensions $d_{\al}, d_{\bt}$
and $d_{\gm}$ respectively. For $k=1,..., d_{\al}$ and $l=1,...,d_{\bt}$ 
and $i,j=1,...,d_{\gm}$ the projection of a tensor product state onto a
state in a single irrep is given by the following 
\begin{defn}
An intertwining mapping between the representations
$(A,\al) \otimes (B,\bt)$ and $(C,\gm)$, 
\be
\rho^C_{\gm,k,l,j}\,:\,L^2_{\al\bt}(G\times G,V_{\al}\otimes V_{\bt})
\to L^2_{\gm}(G,V_{\gm})
\ee
is given by
\bea
\left(\rho^{C(\xi)}_{\gm,k,l,j}\Phi\right)_i (x) &:=& \int_{N_C} \gm_{ij}(h)
\, \Phi_{kl}(xh w(\xi)^{-1}, xh w(\xi)^{-1} y(\xi))\,dh,
\label{eq:rhoC}
\eea
where $C(\xi), w(\xi)$ and $y(\xi)$  are by definition related by
\be
g_A \,\,y(\xi)\, g_B\, y(\xi)^{-1} = w(\xi)\, g_{C(\xi)}\, w(\xi)^{-1}.
\label{eq:xyz}
\ee
\end{defn}
Let us examine the occurring variables more closely: $\xi$ denotes an
element of the double coset $N_A\backslash G /N_B$, which 
parametrises the possible relative difference $y(\xi) = \xinv_1 x_2$
between the entries $(x_1, x_2)$  of $\Phi$. 
Effectively, we have constructed a new function $(\rho \Phi)$ on $G$ out
of a function $\Phi$ on $G\times G$. The new function depends on the
degrees of freedom of the aforementioned relative difference, in the sense
that the conjugacy class $C$ which labels the irrep the new function
belongs to, is uniquely determined by $\xi$ via Eq.(\ref{eq:xyz}). The
variable $y(\xi)$ thus assures that we include the comultiplication
correctly. We assume there exists a continuous mapping 
\be
\xi \mapsto y(\xi)\; : \; N_A\backslash G/N_B \to G,
\label{eq:yxi}
\ee
which via Eq.(\ref{eq:xyz}) determines $w(\xi) \in G$ upto right 
multiplication by an element
$h'\in N_{C(\xi)}$. This assumption does not seem to be restrictive 
\cite{KBM}, and in fact only comes down to fixing the $N_A\otimes N_B$
`phase' factor of the ingoing state $\ABPhi_{\al,\bt}$. By this we mean
that we choose for which $(x_1, x_2) \in G\otimes G$ the $N_A\otimes
N_B$-factor that we can pull out by the covariance condition in 
Eq.(\ref{eq:tensprodspace}) is equal to $\al(e)\otimes \bt(e) = \unit
\otimes \unit$. The choice of the $N_{C(\xi)}$-factor $h'$ on the right of
$w(\xi)$ (in other words: the choice of  
the representative of the coset $w(\xi) N_{C(\xi)}$) is nothing but the
choice of the $N_C$ phase factor of the outgoing function $(\rho\Phi)$. 

If we write $\Phi = |(A,\al)\,a\rangle
|(B,\bt)\, b\rangle = |(A,\al)\,a\,;\,(B,\bt)\, b\rangle$, then
Eq.(\ref{eq:rhoC}) can be rewritten in the more 
familiar form 
\be
\rho^C_{\gm}\left(|(A,\al)\,a\,;\,(B,\bt)\,b\rangle\right) =
\left\{\langle (C,\gm)\,c\,|\,(A,\al)\,a\,;\,(B,\bt)\,b\rangle\right\}
\,\,\,|(C,\gm)\,c\rangle. 
\ee
If the kets span orthogonal bases in the corresponding representation
spaces, we may call the coefficients on the right hand side the
Clebsch--Gordan coefficients of $\DG$. The explicit example of $\DS$ will
be treated in the next section, Clebsch--Gordan coefficients are given in
Eq.(\ref{eq:clebgor}). 

As a last remark on the general formula in Eq.(\ref{eq:rhoC}) we note that
the mapping $\rho^C_{\gm}$ will not exist for all $\gm \in \widehat{N}_C$,
in other words, there are selection rules. 
In \cite{KBM} it was shown that only those $\gm$ are allowed for which
\be
\gm(z) = \al(z)\otimes\bt(z), \quad z\in Z(G),
\label{eq:gamalbet}
\ee
where $Z(G)$ denotes the centre of $G$. 

\subsection*{Decomposition of $\DS$ tensor products}
For the case $G=SU(2)$ the definitions of the previous
paragraph can be made explicit, and the formulas simplify considerably. For
convenience we recall the definition of the Clebsch--Gordan coefficients of
$SU(2)$, using the notation of \cite{VMK}
\be
\De (x) \Dt (x) = \sum_{j=|j_1-j_2|}^{j_1 + j_2} \sum_{m,n=-j}^{j}
C^{jm}_{j_1 m_1 j_2 m_2}\, C^{jn}_{j_1 n_1 j_2 n_2}\,D^j_{mn}(x),\quad x\in G. 
\label{eq:clgordef}
\ee
The Clebsch--Gordan coefficients are the elements of the unitary matrix which
performs the direct and inverse transformations between tensor product state 
vectors $|j_1 m_1 j_2 m_2\rangle= |j_1 m_1\rangle | j_2 m_2\rangle$ and 
$|j_1 j_2 j m \rangle$. The latter is a state in the irreducible 
representation $j$,
coming from the irreducible representations $j_1$ and $j_2$. Since
$C^{jm}_{j_1 m_1 j_2 m_2}=0$ if $m\neq m_1 + m_2$ we can work out the
summations over $m$ and $n$, so for fixed $m_1, m_2, n_1, n_2$:
\be
\De (x) \Dt (x) = \sum'_j C^{j (m_1+m_2)}_{j_1 m_1 j_2 m_2} C^{j 
(n_1+n_2)}_{j_1 n_1 j_2 n_2} D^j_{(m_1+m_2) (n_1+n_2)}(x).
\label{eq:sumjprime}
\ee 
The primed summation runs from max($|j_1-j_2|, |m_1+m_2|,|n_1+n_2|$) to
$(j_1+j_2)$. 
In the following we will work out the summations as far as possible, and
leave out the range of summation for $j$-type labels. 
\vspace*{.2cm}

Let us identify the ingredients for the mapping in Eq.(\ref{eq:rhoC}).
From section \ref{s4} we know that the conjugacy classes $A,B,C$  
for $G=SU(2)$ are denoted by $r_1, r_2, r_3$. The
centraliser representations are labeled by $n_i$ for $r_i\neq 0,2\pi$ and
by $j_i$ otherwise, with dimensions 1 and $(2j_i +1)$
respectively. We now consider various fusion processes, by combining
different sectors:
\vspace*{.2cm}\\
{\bf (I)} The case $r_1, r_2 \neq 0,2\pi$.\\
Consider the tensor product representation
$(r_1, n_1)\otimes (r_2, n_2)$. Take as a basis function 
\bea\label{eq:tensorbasis}
\rrPhi = \reD^{j_1}_{m_1 n_1}\otimes \rtD^{j_2}_{m_2 n_2}\,:\,(x_1,x_2)\mapsto
 \reD^{j_1}_{m_1 n_1}(x_1)\,\rtD^{j_2}_{m_2 n_2}(x_2),\\
\quad j_i\geq n_i,\,-j_i\leq m_i\leq j_i, \,i=1,2.\nn
\eea
We emphasise that $r_1$ and $r_2$ have been fixed. 
In the parametrisation of $SU(2)$ as given in 
Eq.(\ref{eq:gagag}) relation (\ref{eq:xyz}) reads
\be
g_1(r_1,0,0)\,g_2(r_2,\tht,0) = g_3(r_3,\tht_3,\phi_3).
\label{eq:explxyz}
\ee
This means that for $\rrPhi$ we have fixed the $U(1)\otimes U(1)$
phase factor to be 1.
The double coset $U(1)\backslash SU(2)/U(1)$ is parametrised by the
angle $\tht$, and the mapping (\ref{eq:yxi}) is given by
\be
\tht \mapsto y(\tht) = \left(\begin{array}{cc} \cos\half\tht & -\sin\half\tht\\
 \sin\half\tht & \cos\half\tht \end{array}\right).
\ee
Taking the trace on both sides of Eq.(\ref{eq:explxyz}) tells us how
the outgoing conjugacy class label $r_3$ depends on $\tht$:
\be
\cos\frac{r_3(\tht)}{2} = \cos \frac{r_1}{2}\cos \frac{r_2}{2} - \cos\tht
\sin \frac{r_1}{2} \sin \frac{r_2}{2}.
\label{eq:rs}
\ee
From this we see that 
\be
\cos(\frac{r_1+r_2}{2}) \leq \cos\frac{r_3}{2} \leq \cos (\frac{r_1-r_2}{2}),
\label{eq:rrange}
\ee
and thus
\be
|r_1 - r_2|\leq r_3 \leq \mbox{min}(r_1+r_2, 4\pi - (r_1+r_2)),
\ee
which restricts the possible outgoing irreducible representations, i.e.\
the ones in the direct sum/integral in Eq.(\ref{eq:fusion}). In
principle, for all incoming generic representations, one always has 
that $0\leq r_3 \leq 2\pi$, however, in \cite{KBM} it has been shown that
$r_3 = 0$ and $r_3=2\pi$ have measure zero in the image of the mapping
from $U(1)\backslash SU(2)/U(1)$ to $\Conj(SU(2))$. This means that on a
formal level 
special representations do occur in the tensor product decomposition of two
generic representation (f.i.\ when $r_1=r_2$), but that they do {\sl not
contribute} to the squared norm of a generic tensor product state. So in
practice, given two generic representations $(r_1,n_1),\,(r_2,n_2)$, one
doesn't have to compute the mapping $\rho^{r_3}_{n_3}$ for $r_3=0, 2\pi$.

As follows from the  selection rules not all
$n_3\in\hat{N}_{r_3}$ will occur, but only 
\be
n_3 \in \left(\hat{N}_{r_3}\right)_{\ep} = \{ n \in \hat{N}_{r_3} \,|\,
n|_Z = \ep\,\unit \}
\ee
with  
\be
n_1(z)\otimes n_2(z) = \ep (z)\,\mbox{id}_{V_{n_1}\otimes
V_{n_2}},\quad z \in\{e,-e\}\subset SU(2),
\ee
which implies that $n_3$ can only be integer, or half-integer, when
$(n_1+n_2)$ is integer, or half-integer, respectively. Thus the
Clebsch--Gordan series reads 
\be
(r_1,n_1)\otimes (r_2,n_2) \simeq \bigoplus_{n_3\in
(n_1+n_2)\mbox{mod} 
\dubbelZ} \int^{\oplus}_{I_{r_1,r_2}} (r_3,n_3)\, dr_3,
\label{eq:suseries}
\ee
where $I_{r_1,r_2} = [|r_1-r_2|, \mbox{min}(r_1+r_2, 4\pi-r_1-r_2)]$.

The variables $w(\tht)$ and $y(\tht)$ are related via 
\be
g_{r_1} y(\tht) g_{r_2} y(\tht)^{-1} = w(\tht) g_{r_3} w(\tht)^{-1}.
\label{eq:xyzsu2}
\ee
From Eq.(\ref{eq:explxyz}) we can also derive that
\be
\sin\frac{r_3}{2} \hat{n}_3 = \sin\frac{r_1}{2}\cos\frac{r_2}{2}\,\hat{n}_1
+ \cos\frac{r_1}{2}\sin\frac{r_2}{2}\,\hat{n}_2 - \sin\frac{r_1}{2}
\sin\frac{r_2}{2}\,(\hat{n}_1\wedge \hat{n}_2)
\ee
where $\hat{n}_1\wedge \hat{n}_2$ denotes the exterior product, and
$\hat{n}_3$ corresponds to $w(\tht)$, see Eqs.(\ref{eq:gagag}) and
(\ref{eq:exponent}). The choices made in Eq.(\ref{eq:explxyz}) mean that 
\bea
\hat{n}_1 = \left(\begin{array}{c} 0 \\ 0 \\ 1\end{array}\right), \quad
\hat{n}_2 = \left(\begin{array}{c} \sin\tht \\ 0 \\ \cos\tht\end{array}\right)
\eea
and thus we can calculate 
\bea
\hat{n}_3 = \hat{n}_{w(\tht)} =  \left(\begin{array}{c} 
\sin \tht_w \cos\phi_w\\ \sin \tht_w \sin\phi_w\\ \cos \tht_w 
\end{array}\right) =  
\frac{1}{\sin\frac{r_3}{2}}  \left(\begin{array}{c}
\sin\tht\cos\frac{r_1}{2} \sin\frac{r_2}{2}\\ 
\sin\tht \sin\frac{r_1}{2} \sin\frac{r_2}{2} \\ 
\sin\frac{r_1}{2} \cos\frac{r_2}{2} +\cos\tht\cos\frac{r_1}{2}
\sin\frac{r_2}{2}\end{array}\right).
\eea

The irreducible unitary representations $n_1, n_2, n_3$ of $N_{r_1}=
N_{r_2}=N_{r_3}=U(1)$  are 1-dimensional, so the
indices $i,j,k,l$ in Eq.(\ref{eq:rhoC}) won't occur. 
The mapping $\rho$ on $\rrPhi$ now reads:
\bea
\left(\rho^{r_3(\tht)}_{n_3} \Phi\right)(x) &=& \int_{U(1)} e^{in_3\zeta}
\left(\rrPhi(xhw(\tht)^{-1},xhw(\tht)^{-1}y(\tht))\right)\, dh \nn\\
&=& \sum_{j=|j_1-j_2|}^{j_1+j_2}\sum_{m,p=-j}^{j} 
\sum_{p_2=-j_2}^{j_2}  C^{jm}_{j_1 m_1 j_2 m_2} C^{jp}_{j_1 n_1 j_2 p_2}
D^{j_2}_{p_2 n_2}(y(\tht))\,\times \nn\\
& &\qquad\qquad\int_{U(1)} e^{in_3\zeta} 
\sum_{r,s=-j}^{j} D^j_{mr}(x) D^j_{rs}(h) D^j_{sp}(w(\tht)^{-1})\,dh \\
&=:& \sum'_j 
\langle (r_3,n_3) j m\,|\,(r_1,n_1) j_1 m_1; (r_2,n_2) j_2 m_2\rangle\,\,
D^j_{m n_3}(x),\nn
\label{eq:su2decom}
\eea
with $h=e^{i\zeta}$. The primed summation over $j$ has been explained
under Eq.(\ref{eq:sumjprime}), the summation over $p_2$ runs from
max($(-j-n_1), -j_2)$ to min($(j-n_1), j_2)$. From this it follows that the
Clebsch--Gordan coefficients of the quantum double group of $SU(2)$ are
given by 
\be
\langle (r_3,n_3) j m\,|\,(r_1,n_1) j_1 m_1; (r_2,n_2) j_2 m_2\rangle = 
\sum'_{p_2} C^{j m}_{j_1 m_1 j_2 m_2} 
C^{j (n_1+p_2)}_{j_1 n_1 j_2 p_2}  \,
D^{j_2}_{p_2 n_2} (y(\tht)) \overline{D^j_{(n_1+p_2) n_3}(w(\tht))}.
\label{eq:clebgor}
\ee
Clearly they depend on the representation labels $(r_i,n_i), i=1,2,3$
and on the specific states the functions represent, which are labeled by 
the $j_1,m_1$, etc., just as would be expected. 
\vspace*{.2cm}\\
{\bf (II)} The case $r_1=r_2=0$\\
In case both incoming `magnetic fluxes' are trivial (the conjugacy
classes of the corresponding irreducible representations are 
the unit element in the group) the `electric charges' (centraliser 
representations)  correspond to 
representations of the full group $SU(2)$, we will denote them
by $j_1$ and $j_2$ respectively. The constraining relation (\ref{eq:xyzsu2}) 
shows that now only $r_3=0$ can occur, and that $y(\tht)$ and $w(\tht)$
never affect this constraining relation. Any $\tht$ (and thus $y(\tht)$
and $w(\tht)$) will satisfy Eq.(\ref{eq:xyzsu2}), so in the definition of
$\rho$ 
we may simply take them to be unity. This means that we take the
same basis relative to which we choose the basis in the incoming
$SU(2)$-fibres. So, for these special 
representations we can in fact consider the tensor product states as
sections in the same point.  

The $(m_1, m_2)$-component of the incoming two particle wavefunction reads
\be
\left(\Phi_{n_1,n_2}\right)_{m_1,m_2}(x_1, x_2) = (\De \otimes \Dt)(\xinv_1, 
\xinv_2). 
\ee
The $n_1$ and $n_2$ occur for the same reason as for the single irreducible
representation functions, as explained under Eq.(\ref{eq:hilbspaces}), i.e.\
they are related to the basis vectors $v_{n_1}\otimes v_{n_2}$ in 
$V_{j_1}\otimes V_{j_2}$. The mapping $\rho$ reads
\bea
\lefteqn{\left(\rho^0_{j_3, m_1,m_2,n_3} \Phi\right)_{m_3}(x) = \int_{SU(2)}
\Dd(g) \De(\ginv \xinv) \Dt(\ginv \xinv)\,dg = \quad} \nn\\
&=& \sum_{p_1= -j_1}^{j_1}\sum_{p_2=-j_2}^{j_2} \left(\int 
\overline{D^{j_3}_{n_3
m_3}(\ginv)}\,D^{j_1}_{m_1 p_1}(\ginv) D^{j_2}_{m_2 p_2}(\ginv)\,dg\right)
D^{j_1}_{p_1 n_1}(\xinv) D^{j_2}_{p_2 n_2}(\xinv) = \nn\\
&=& \frac{1}{2j_3+1} C^{j_3 n_3}_{j_1 m_1 j_2 m_2}
C^{j_3 (n_1+n_2)}_{j_1 n_1 j_2 n_2}\, D^{j_3}_{m_3 (n_1+n_2)} (\xinv).
\label{eq:rho000}
\eea
From the properties of the $SU(2)$ Clebsch--Gordan coefficients we see that 
this mapping only differs from zero if $n_3=m_1+m_2$. Eq.(\ref{eq:rho000}) 
gives the $m_3$-rd component of a section which is related to $v_{n_1+n_2}
\in V_{j_3}=\CC^{2j_3+1}$ as indicated under Eq.(\ref{eq:hilbspaces}). 

We have combined two pure electric states corresponding to 
irreducible unitary $SU(2)$-representations, and decomposed it in
single pure electric states. This should be completely equivalent to
the case of ordinary $SU(2)$-representation theory, so we would like to make
explicit connection with that. To that aim recall our discussion at the
end of section \ref{s4}, where we remarked that in order to obtain a
normal $SU(2)$-representation one should fix the label $n$ of the functions
$D^j_{mn}$. The same holds for tensor products, that is, one should fix
$n_1$ and $n_2$, which is nothing more than choosing a specific basis in the
tensor product representation space. This implies that the basis in the
irreducible components has been fixed  to ($n_1+n_2$). The second
$SU(2)$-Clebsch--Gordan coefficient in Eq.(\ref{eq:rho000}) now is only a
fixed number, depending on the basis choice. Indeed only one
$SU(2)$-Clebsch--Gordan coefficient remains, and we have recovered the
ordinary $SU(2)$ representation theory. 
{\bf (III)} The case $r_1\neq 0,2\pi,\,r_2=0$\\
We consider the tensor product of a generic representation with a special
representation corresponding to a flux $e$. By construction we can never
have that $r_3=0,2\pi$. The 
incoming wavefunction takes values in $\CC \otimes V_{j_2} \simeq
V_{j_2}$, and we take the $m_2$-component to be given by 
$\left(\Phi(z_1, z_2)
\right)_{m_2} = \De(z_1) \Dt(\zinv_2)$. Relation (\ref{eq:xyzsu2}) tells us
that $r_3 = r_1$ and that $w(\tht)\in N_{r_3}=U(1)$. As under ({\bf
II}) the variable $y(\tht)$ may take any value. From
Eq.(\ref{eq:xyzsu2}) it follows that $w(\tht)=e$. Thus we get
\bea
\lefteqn{\left(\rho^{r_3}_{n_3,m_2}\Phi\right)(x) = \int_{U(1)} 
e^{in_3\zeta}\,\De(xh w(\tht)^{-1})\,\Dt((xh w(\tht)^{-1})^{-1})\,dh=
\qquad\quad}\nn\\
&=& \int_{U(1)} e^{in_3\zeta}D^{j_1}_{m_1 n_1}(x)D^{j_2}_{m_2 n_2}(\xinv)
e^{-in_1\zeta} e^{im_2\zeta}\,d\zeta =\nn\\
&=& \sum'_j \sum_{m,n=-j}^j C^{jm}_{j_1 m_1 j_2 -n_2}\,
C^{jn}_{j_1 n_1 j_2 -m_2} (-1)^{n_2-m_2} \dl_{n_3,n_1-m_2} D^j_{mn}(x)=\nn\\
&=&\sum'_j \{(-1)^{n_2-m_2} C^{j (m_1-n_2)}_{j_1 m_1 j_2
-n_2}\,C^{j (n_1-m_2)}_{j_1 n_1 j_2 -m_2} \,\dl_{n_3,n_1-m_2}\}\,
D^j_{(m_1-n_2)n_3}(x).
\label{eq:rhor10}
\eea
As a first check we see that this function has the correct behaviour under
right multiplication of its argument by an arbitrary element of $U(1)$. 
As  expected, the outgoing electric charge $n_3$ depends on the
$SU(2)$-state of the second incoming particle. 
This $m_2$ can vary between $-j_2$ and $j_2$, so the
outgoing electric label $n_3$ can vary between $(n_1-j_2)$ and
$(n_1+j_2)$, if we don't specify the electric state of the second
ingoing particle. The corresponding Clebsch--Gordan coefficient reads
\be
\langle (r_3,n_3) j m\,|\,(r_1,n_1) j_1 m_1; (0,j_2) m_2\rangle = 
(-1)^{m_1-m_2-m} C^{j m}_{j_1 m_1 j_2
(m-m_1)}\,C^{j (n_1-m_2)}_{j_1 n_1 j_2 -m_2} \,\dl_{n_3,n_1-m_2},
\ee
as follows from Eq.(\ref{eq:rhor10}).\\
{\bf (IVi)} $r_1\neq 0,2\pi$ and $r_2=2\pi$. Same result as
Eq.(\ref{eq:rhor10}),  
only with $r_3=r_1+2\pi$, due to the fact that Eq.(\ref{eq:xyzsu2})
now reads $g_{r_1} (-e) = -g_{r_1} = g_{r_1+2\pi} = w g_{r_3} \winv$,
where again we may choose $w=e$. \\
{\bf (IVii)} $r_1=r_2= 2\pi$. Same as Eq.(\ref{eq:rho000}), with $r_3=0$. 
\section{The action of the $R$-element}\label{s6}
\setcounter{equation}{0}
The existence of the $R$-element in ${\cal D}(SU(2))$ enables us to
compute the action of the braid group on irreducible unitary
representations. The action of the universal $R$-element reads
\be
\left( \left((A,\al) \otimes (B,\bt)\right)R\right)\,\Phi(x, y) =
  \Phi(x, x g_A^{-1}\xinv y).
\ee
The braid operator $\cal{R}$ is an intertwining mapping between
$(A,\al) \otimes (B,\bt)$ on $V_{\al}\otimes V_{\bt}$ and $(B,\bt)\otimes
(A,\al)$ on $V_{\bt}\otimes V_{\al}$ given by
\be
{\cal{R}}^{AB}_{\al\bt}\,\Phi := \left(\sigma_L \circ \left((A,\al) \otimes
(B,\bt) \right)(R)\right) \Phi 
\ee
where
\be
(\sigma_L \Phi)(x,y) = \sigma \left(\Phi(y,x)\right),\quad 
\sg(v\otimes w):= w\otimes v,\quad v\in V_{\al}, w\in V_{\bt},
\ee
interchanging the representations $(A,\al)$ and $(B,\bt)$. Hence
\be
\left({\cal{R}}^{AB}_{\al\bt}\,\Phi\right)(x,y) =
\sigma_L\left(\left((A,\al) \otimes (B,\bt) \right)(R) \Phi(x,y)\right) =
\sigma \left(\Phi(y, y g_A^{-1}\yinv x)\right). 
\label{eq:rab}
\ee
At first sight, comparing to Eq.(\ref{eq:quantumbraiding}), this is not
quite what one would expect. However, a state $|g\rangle$ corresponds to 
the delta function of Eq.(\ref{eq:deltasections}), and it 
can be checked that by taking $\Phi = \dl_{x_1}\otimes
\dl_{x_2}$ Eq.(\ref{eq:rab}) leads to the correct interchange
transformation. This has been done for ${\cal R}^2$ in
Eq.(\ref{eq:deltascattering}).   

With the previous equation it is not difficult to show that 
\be
({\cal R}^2 \Phi)(y_1,y_2) = \Phi (y_2 \ginv_B \yinv_2 y_1, 
y_2 \ginv_B \yinv_2 y_1 \ginv_A \yinv_1 y_2 g_B \yinv_2 y_2),
\label{eq:R2}
\ee
where ${\cal R}^2 = {\cal R}^{BA}_{\bt\al} {\cal R}^{AB}_{\al \bt}$. 
For our $SU(2)$ example we can simply choose a basis function for
$\Phi$.  The monodromy operation on such a function,
i.e.\ winding one of the particles counterclockwise around the other, is the
action of ${\cal R}^2$ in the representation $(r_1,n_1)\otimes
(r_2,n_2)$, and it reads
\be
{\cal R}^2 \left(D^{j_1}_{m_1 n_1}\otimes D^{j_2}_{m_2 n_2}\right)(y_1,
y_2) =  \left(D^{j_1}_{m_1 n_1}\otimes D^{j_2}_{m_2 n_2}\right)(y_2
\ginv_{r_2} \yinv_2 y_1, 
y_2 \ginv_{r_2} \yinv_2 y_1 \ginv_{r_1} \yinv_1 y_2 g_{r_2}).
\label{eq:R2act}
\ee

As alluded to before, our Hilbert space must also form a unitary
irreducible representation of the pure braid group. For two particles this
group $B_2(\Sg)$ is abelian, so it has one-dimensional irreps. The
eigenvalue of the monodromy operator must satisfy the 
generalised spin--statistics connection \cite{DPR}
\be
K^{ABC}_{\al\bt\gm} {\cal R}^2 = e^{2\pi i(s_{(C,\gm)} - s_{(A,\al)} - 
s_{(B,\bt)})} K^{ABC}_{\al\bt\gm}
\label{eq:spinstat}
\ee
where $K^{ABC}_{\al\bt\gm}$ denotes the projection on the irreducible
component 
$(C,\gm)$ of $(A,\al)\otimes (B,\bt)$, given by our mapping $\rho^C_{\gm}$ in 
Eq.(\ref{eq:rhoC}). The
(generalised) spin $s_{(A,\al)}$ of the sector $(A,\al)$ is given by the
action of the central element of the quantum double on a state in the
$(A,\al)$-representation:
\be
\left( \left(\int_G dg\, \dl_g\otimes g\right)\, \aphi_{\al}\right) (y)
= \aphi_{\al}(y \ginv_A) = \al(g_A)\, \aphi_{\al}(y) = e^{2\pi i
s_{(A,\al)}}\,\aphi_{\al}(y).
\ee
For the case $G=SU(2)$ this means that 
\be
e^{2\pi i s_{(r,n)}} = e^{inr}\quad\longleftrightarrow\quad s_{(r,n)} = 
\frac{nr}{2\pi}.
\label{eq:spin}
\ee
We also verify Eq.(\ref{eq:spinstat}) for this case
\bea
\lefteqn{\left(\rho^{r_3}_{n_3} {\cal R}^2 \left(D^{j_1}_{m_1 n_1}\otimes
D^{j_2}_{m_2 n_2}\right) \right)(x)= \qquad}\nn\\
& &\qquad = \int_{U(1)} dh\, n_3(h)\, D^{j_1}_{m_1 n_1}(xh\ginv_{r_3} \winv
g_{r_1}, xh\ginv_{r_3} \winv y g_{r_2} ) = \nn \\
& &\qquad = e^{-in_1 r_1} e^{-in_2 r_2}\,\int_{U(1)} dh\,n_3(h
g_{r_3}) \,(D^{j_1}_{m_1 n_1}\otimes D^{j_2}_{m_2 n_2})(xh\winv, 
xh\winv y) = \nn\\
& &\qquad = e^{-in_1 r_1} e^{-in_2 r_2} e^{in_3 r_3}\left(\rho^{r_3}_{n_3} 
(D^{j_1}_{m_1 n_1}\otimes D^{j_2}_{m_2 n_2})\right) (x),
\eea
where we have used that $g_{r_1} y g_{r_2} \yinv = w g_{r_3} \winv$. 
\subsection{Scattering}
The fact that the internal quantum numbers of encircling point
particles with purely topological interactions affect the scattering
cross--section is well known from the Aharonov--Bohm effect
\cite{AhBo}, which deals with charged point particles with fluxes
exponentiating to elements of an abelian group. We recall 
the differential scattering cross--section for an external charge
$q$ (corresponding to an irreducible representation of $U(1)$) 
scattering off a particle with only magnetic flux $\phi$ (corresponding
to the $U(1)$-element $e^{i\phi}$):
\be
\frac{d\sg}{d\tht} = \frac{1}{2\pi p}
\frac{\sin^2\frac{q\phi}{2}}{\sin^2 \frac{\tht}{2}}
\ee
where $p$ denotes the momentum of the incoming particle,
%(better, plane wave, because we are dealing with quantum mechanics), 
and $\tht$ is the scattering angle. The nonabelian generalisation
of this formula involves the monodromy ${\cal R}^2$, which now typically
will be nondiagonal, unless we have chosen a particular incoming
two-particle state being an eigenstate of the monodromy operator. It
has been derived by Verlinde \cite{Verlinde}
\be
\frac{d\sg}{d\tht} = \frac{1}{4\pi p\sin^2 \frac{\tht}{2}}(1 - \mbox{Re}
\langle\mbox{in}|{\cal R}^2|\mbox{in}\rangle ).
\label{eq:crossec}
\ee
As explained in \cite{WPB}, this formula was obtained by
understanding that the braid group for the two-particle case is
abelian, and therefore the monodromy matrix ${\cal R}^2$ can be
diagonalised, reducing the nonabelian problem to a --solved-- abelian
problem. It is also important to remark that this is an {\sl
inclusive} cross--section, which means that the detector measuring the
scattered particles does not distinguish between their internal electric
states, in other words, scattering may go via all possible fusion
channels. This is made more clear by Eq.(\ref{eq:sumchannels}). 

In general, a two-particle Hilbert space can be decomposed into
representation spaces of quantum double irreps, as shown in  the former
section. This means that a complete basis for this two-particle Hilbert
space can be given by the direct sum of bases of all possible irreps of the
quantum double. If we denote this basis by $\{|(A,\al)\,a\rangle\},\,\,A\in
\Conj(G), \al\in\hat{N}_A$, and $a$ labeling the basis functions on
$C_A$ as in Eq.(\ref{eq:braket}), we can write
\be
\langle \,\mbox{in}\, |\,{\cal R}^2\, |\,\mbox{in}\,\rangle = \int_{\Conj(G)}
d\mu(A)\,\sum_{\al, a}\,\langle\, \mbox{in}\,| (A,\al)\,a\rangle \, \langle
(A,\al)\,a\, |\,{\cal R}^2\, |\, \mbox{in}\,\rangle.
\label{eq:compl}
\ee
To be more specific, for $G=SU(2)$ the complete basis $\{|(A,\al)\,a
\rangle\}$ is given by 
\bea
\left\{ \, |(r_3,n_3) j m\rangle,\;\mbox{with}\; 0< r_3 < 2\pi, n_3
\in \half\dubbelZ, j\geq n_3, j\in \half\dubbelN, -j\leq m\leq j \right\}\;\;
\bigcup \nn\\
 \left\{ \, |(0,l) m'\rangle, |(2\pi, l) m'\rangle, \; \mbox{with}\;\;
l\in \half\dubbelN, -l\leq m'\leq l \right\}.
\eea
We will now calculate the differential cross--sections for the
scattering of various types of particles.
\vspace*{.2cm}\\
(i) Consider two generic representations, so the incoming
state is an element of $V^{r_1}_{n_1}\otimes V^{r_2}_{n_2}$. \\
Firstly, take the incoming state $|$in$\rangle$ as given in
Eq.(\ref{eq:tensorbasis}).  Then
the first factor in Eq.(\ref{eq:compl}) is 
simply the Clebsch--Gordan coefficient from Eq.(\ref{eq:clebgor}). We
have seen that these Clebsch--Gordan coefficients are only nonzero for
a subset of all possible fusion channels, so the sum/integral in
Eq.(\ref{eq:compl}) is only over the irreps that occur in the
Clebsch--Gordan series, see Eq.(\ref{eq:suseries}).
For the second factor of Eq.(\ref{eq:compl}) we use the
generalised spin--statistics connection of Eq.(\ref{eq:spinstat})
\be
\langle (r_3,n_3) jm\,|\,{\cal R}^2\, |\,\mbox{in}\,\rangle = e^{i(n_3 r_3
-n_1 r_1 -n_2 r_2)} \langle  (r_3,n_3) jm\, |\,\mbox{in}\,\rangle. 
\label{eq:channelspin}
\ee
We arrive at
\bea
\lefteqn{
\langle\,\mbox{in}\, |\,{\cal R}^2\, |\, \mbox{in}\,\rangle = e^{-i(n_1 r_1+n_2
r_2)}\!\!\!\!\!\!\!\!\!
\sum_{n_3\in (n_1+n_2)\mbox{\scriptsize{mod}\,$\dubbelZ$}}\,\,\sum'_{j,m}
\hfill} \nn\\ 
& &\qquad\qquad\qquad\int_{I_{r_1,r_2}} 
e^{in_3 r_3} |\langle (r_1,n_1) j_1 m_1, (r_2,n_2) j_2 m_2\,|\,
(r_3,n_3) j m\rangle |^2\; dr_3.
\label{eq:sumchannels}
\eea
Unfortunately, it cannot easily be seen that this expression is
finite, which we expect to be the case.  An alternative way is to compute
the left hand side of Eq.(\ref{eq:channelspin}) directly. We use the
explicit action of ${\cal R}^2$ as given in Eq.(\ref{eq:R2act}), and
take the inner product according to Eq.(\ref{eq:tensinprod}). Working
it all out leaves us with the following unattractive, but finite (!),
expression:
\bea \label{eq:bigsum}
\lefteqn{ \langle\mbox{in} |{\cal R}^2 | \mbox{in}\rangle =\!\!
\sum_{j=|j_1-j_2|}^{j_1+j_2} \sum_{p_2,q_2,s_2,u_2=-j_2}^{j_2}
\sum_{m,m',p,q=-j}^{j} \sum_{k,l=|j-j_2|}^{j+j_2} \sum_{\kp,
\kp',\lm,\lm'=-k,l}^{k,l} \sum_{b=|k-j_2|}^{k+j_2}
\sum_{\bt,\bt'=-b}^{b} \qquad } \nn\\
& & e^{-ip_2\frac{r_1}{2}} e^{-i(p-s_2)\frac{r_2}{2}}\,\,
C^{jm}_{j_1 m_1 m_2 m_2}\,C^{j m'}_{j_1 n_1 j_2 p_2}\, C^{j
m'}_{j_1 n_1 j_2 p_2}\, C^{k\kp}_{j_2 q_2 j m}\, C^{k\kp'}_{j_2 n_2 j
p}\, C^{l\lm}_{j_2 m_2 j q}\, C^{l\lm'}_{j_2 n_2 j p}\, C^{jn}_{j_1
m_1 j_2 u_2} \,\times \nn\\
& &\qquad \qquad \qquad \qquad\qquad C^{b\bt}_{k\kp j_2 u_2}\,
C^{b\bt'}_{k\kp' j_2 s_2} \, C^{b\bt}_{l \lm j_2 q_2}\, C^{b\bt'}_{l\lm'
j_2 s_2} \nn\\ 
\eea
Inserting this in Eq.(\ref{eq:crossec}) yields a finite differential
scattering cross--section for the case where the incoming wave
function is given by $\reD^{j_1}_{m_1 n_1}\otimes \rtD^{j_2}_{m_2
n_2}$.

Secondly, as a different example of an incoming state in the same
representation, we consider a wave function corresponding to
two particles with sharply determined fluxes $g_1 = x_1 g_{r_1}
\xinv_1$ and $g_2 = x_2 g_{r_2} \xinv_2$. This means that $|$in$\rangle$
is given by ${}^{r_1}\!\dl_{x_1, n_1}\otimes {}^{r_2}\!\dl_{x_2,n_2}$,
a tensor product of `delta-sections' for which
\be
\dl_{x\hinv, n} = e^{-in\zeta}\dl_{x,n},\quad \forall h=e^{i\zeta}\in
U(1),
\ee
where from now on we omit the $r$-labels, as before. 
Formally, such delta-sections are not
elements of the Hilbert spaces of irreps of ${\cal D}(SU(2))$, because
they are not square integrable (normalisable), as was required in
Eq.(\ref{eq:hilbertspace}). However, they correspond to the more
classical picture of incoming particles of which we know the fluxes
precisely, therefore we would like to know the scattering
cross--section for such wave functions (distributional sections) as
well. From the model for topologically interacting excitations in
broken gauge theories \cite{WPB}, which corresponds to the quantum double
of a finite group, we know that the action of the monodromy operator
comes down to conjugation of each flux by the total flux, and that there
can be nontrivial actions in the centraliser representation spaces. 
%\be
%{\cal R}^2\;:\;|g_1\rangle |g_2\rangle\;\mapsto |(g_1 g_2) g_1 (g_1
%g_2)^{-1} \rangle | g_1 g_2 \ginv_1\rangle. 
%\ee
It is easy to check that this corresponds to the action of ${\cal
R}^2$ as given in Eq.(\ref{eq:R2}), which now reads
\be
\left({\cal R}^2 \; \dl_{x_1}\otimes \dl_{x_2}\right)(x,y) = e^{-in_1
r_1}\,\left( \dl_{g_1 g_2 x_1} \otimes \dl_{g_1 x_2}\right)(x,y).
\label{eq:deltascattering}
\ee

Consider the inner product between two delta-sections in a generic
irrep $(r,n)$:
\be
\langle \dl_x\,|\,\dl_{x'}\rangle = \int\,\langle \dl_{x,n}(y)\,,\,
\dl_{x',n}(y)\rangle_{V_n}\, dy = \dl_{x',n}(x).
\ee
We decide that we extract all non-equal $U(1)$ phase factors from the
right. This means the following: in general any $x\in G$ can be
written as $\bar{x}h$, with $\bar{x}\in G$ a chosen representative of the
coset $xH$, and $h\in H$. To  compare differences in centraliser states
between elements of the Hilbert space $V^r_n$ we choose the same
representative in each coset. For the $SU(2)$ case in the Euler--angle
parametrisation this comes down to  choosing the same angle $\psi$, see
Eq.(\ref{eq:euler}). So $x=\bar{x}h$ and $x'=\bar{x}'h'$, with now
$x,x' \in SU(2)$, and $h,h' \in U(1)$. This yields
%Then $h' = h \tilde{h}$ for a unique $\tilde{h}\in U(1)$, and we write 
\be
\dl_{x',n}(x) = n(x'\xinv) \dl_{\bar{x}'}(\bar{x})
\ee
where the delta on the right hand side is an ordinary delta function on the
group. In calculating $\langle\mbox{in} |{\cal R}^2 | \mbox{in}\rangle$ we 
find
\be
\langle\mbox{in} |{\cal R}^2 | \mbox{in}\rangle = e^{-in_1
r_1}\,\dl_{g_1 g_2 x_1,n_1}(x_1)\,\dl_{g_1 x_2,n_2}(x_2).
\ee
This will always be zero, unless $g_1$ and $g_2$ commute. They do so
iff $\bar{x}_1 = \bar{x}_2$, or $\bar{x}_1 = \bar{x}_2
a_{\pi}$. In these cases we can `pull' the factors $g_1 g_2$ and $g_1$
through $x_1$ and $x_2$ respectively. Since we have chosen the
representatives of the conjugacy classes to lie in (the same) $U(1)$
we obtain
\be
\dl_{g_1 g_2 x_1, n_1}(x_1) = e^{in_1(r_1+r_2)} \dl_{x_1}(x_1),
\ee
and a similar factor for $\dl_{x_2, n_2}$, and similar term for the
case  $\bar{x}_1 = \bar{x}_2 a_{\pi}$. Such a fulfilled delta function
of course gives rise to infinities, which are to be expected, since we
started off with non-normalisable states. 
We find that
\be
\langle\mbox{in}\, |{\cal R}^2 |\,\mbox{in}\rangle
= e^{in_1 r_2} e^{in_2
r_1} \dl_{x_1}(x_2) + e^{-in_1 r_2} e^{-in_2 r_1} \dl_{x_1}(x_2
a_{\pi}),
\ee
and upon substituting this in Eq.(\ref{eq:crossec}) we see that the
differential cross section for the scattering of two particles with
magnetic fluxes $g_1$ and $g_2$ (and equal electric phase factor) becomes
\be
\frac{d\sg}{d\tht} = \frac{1}{4\pi p}\frac{1}{\sin^2\frac{\tht}{2}}
\left(1- (\dl_{x_1}(x_2)\, + \dl_{x_1}(x_2 a_{\pi}))\,\cos(n_2 r_1+n_1
r_2) \right).
\ee
Note that the delta functions live on the set of coset
representatives. 

From this we conclude that the scattering of two precisely determined
fluxes that do not commute gives 
\be
\frac{d\sg}{d\tht} = \frac{1}{4\pi p}\frac{1}{\sin^2\frac{\tht}{2}}.
\end{equation}
If the fluxes do commute we in a sense retrieve the abelian case, were
it not for the fulfilled delta functions in front of the cosine. If
these are somehow regularised (by approximating the continuous set of
coset representatives by a finite set, and taking Kronecker delta's
instead of Dirac delta functions) we find that
\be
\frac{d\sg}{d\tht} = \frac{\sin^2\half(n_2 r_1+n_1 r_2)}{2\pi
p}\frac{1}{\sin^2\frac{\tht}{2}}.
\end{equation}
\vspace*{.2cm}\\
(ii) In this case we consider the incoming state to correspond to a generic
and a special representation, $|$in$\rangle \in V^{r_1}_{n_1}\otimes
V^0_{j_2}$. In particular, we take 
\be
|\mbox{in}\rangle = \rDe \otimes {}^0\!\Dt,
\ee
then it can be shown that
\be
\frac{d\sg}{d\tht} = \frac{1}{4\pi p}\frac{1}{\sin^2\frac{\tht}{2}}
\left(1- \cos(r_1 m_2)\, \sum'_j\left(C^{j (m_1-n_2)}_{j_1 m_1 j_2
-n_2}\, C^{j (n_1-m_2)}_{j_1 n_1 j_2 -m_2}\right)^2 \right),
\ee
where we have used Eq.(\ref{eq:rhor10}).
\vspace*{.2cm}\\
(iii) Consider again a generic and a special representation, now with
$|$in$\rangle = \rDe \otimes {}^{2\pi}\!\Dt$. This will give
\be
\frac{d\sg}{d\tht} = \frac{1}{4\pi p}\frac{1}{\sin^2\frac{\tht}{2}}
\left(1+\cos((n_1+m_2)r_1 +2\pi m_2)\,\sum'_j\left(C^{j
(m_1-n_2)}_{j_1 m_1 j_2 
-n_2}\, C^{j (n_1-m_2)}_{j_1 n_1 j_2 -m_2}\right)^2 \right).
\ee
(iv) For pure charges, so $|$in$\rangle \in V^0_{j_1}\otimes
V^0_{j_2}$ it is easy to see that the monodromy matrix leaves the
incoming state invariant, and thus $\frac{d\sg}{d\tht}=0$. 

\section{Gravity in 2+1 dimensions, and ${\cal D}(SO(2,1))$.} 
\label{sgrav}
\setcounter{equation}{0}
In this section we show that, under the assumption of zero cosmological
constant, for gravitating particles in two spatial and one time
dimension there also is the structure of a quantum double symmetry, now 
corresponding to the locally compact group $SO(2,1)$. We briefly
recall the special features of 2+1 gravity, and the classical point
particle solutions. Then we summarise Witten's gauge theoretical
formulation of 2+1 gravity, and show that it is almost
equal to the CS theory discussed in section \ref{s2}, but now for the
gauge group $ISO(2,1)$ (or its universal covering, but for now we only
consider $ISO(2,1)$).  

We start with a general remark about gravity in three dimensions, which
shows the topological nature of the theory. The Einstein--Hilbert action
is given by    
\be
S_{EH} = \int d^3 x \, \sqrt{-g} \, R,
\label{eq:EinsteinHilbert}
\ee
with $g$ the determinant of the full three-dimensional metric of the
spacetime, and $R$ the associated scalar curvature. In general, that is
in  arbitrary dimensions, the equations of
motion are the Einstein equations, $G_{\mu\nu}=R_{\mu\nu}-\half
g_{\mu\nu}R=0$, with $R_{\mu\nu}$ the 
Ricci tensor and $G_{\mu\nu}$ the Einstein tensor. When matter terms 
are added to the action the field equations become
\be
G_{\mu\nu}=\kp T_{\mu\nu}, 
\label{eq:einsteineqs}
\ee
where $T_{\mu\nu}$ is the energy-momentum tensor, $\kp=8\pi G$ the
coupling constant, and $G$ Newton's constant. The Riemann four-index
tensor $R_{\al\mu\bt\nu}$ carries all the information about the
curvature of space time. What makes three dimensional space time
special is that in that case the curvature tensor $R_{\al\mu\bt\nu}$ and
Ricci tensor  $R_{\mu\nu}$ have the same number of degrees of
freedom. This allows one to express the curvature tensor in terms of
the Einstein tensor
\be
{R^{\al\bt}}_{\mu\nu}=-{\ep^{\al}}_{\mu\gm}{\ep^{\bt}}_{\nu\dl}
G^{\gm\dl}.
\ee
Together with Eq.(\ref{eq:einsteineqs}) we immediately see that in
empty regions the curvature tensor vanishes, so the geometry is
flat. We will only consider point sources, so the curvature is
concentrated along the worldlines of the particles. 

\vspace*{.2cm}

Next, we briefly consider classical particle solutions in this
theory \cite{DJH}. From now on we put $8\pi G =1$. For a single,
static, particle of mass $\mu$ located at the origin the energy-momentum
tensor reads 
\be
T^{00}(z)= \mu\, \delta^{(2)}(z), \quad T^{0i}=T^{ij}=0.
\ee
Solving the Einstein equations in the ADM formulation of general
relativity, and making an appropriate coordinate transformation, yields
the following expression for the spatial line element
\be
 dl^{2}= dr^{2} + r^{2}d\theta^{2}, 
\label{eq:line}
\ee
which is manifestly flat. However, the range of the angular variable
$\theta$ is restricted  
\be
 0\leq \theta \leq 2\pi\alpha,\qquad \al:= 1 - \frac{\mu}{2\pi}. 
\ee
For $\mu<2\pi$ this corresponds to a cone with deficit angle $\mu$. 
These boundary conditions are equivalent to the holonomy one obtains
upon performing a closed loop around the particle in the original, not
manifestly flat coordinates. (This can for instance be measured by
carrying a gyroscope around the particle, because the holonomy corresponds
to a rotation over $\mu$.)  Only values $0 \leq \mu <2\pi$ 
are physically admissable, because for masses equal or larger than
$2\pi$ the space vanishes. In case the particle also carries
intrinsic angular momentum, i.e.\ it has spin $s$, it can be shown that
upon performing a closed loop the holonomy also contains a jump in time of
magnitude $s$, in our units. Here we will not discuss the possible
consequenses of the metric described above, but restrict ourselves
to configurations whose total mass is positive, but smaller than $2\pi$,
and allow for translations in time. 

To a mass $\mu$ spin $s$ particle in the origin we thus associate the
holonomy corresponding to the element $(\Lm_{\mu}, a_s)$ of the
Poincar\'e group $ISO(2,1) = SO(2,1) \semdir T(3)$ in three dimensions,
which is a pure rotation over $\mu$ and a pure time translation over
$s$. (From now on we omit vector arrows for three-vectors, to avoid
confusion with spatial two-vectors.) 
In case the particle is not static but moving in an arbitrary Lorentz
frame, an observer should first boost (with an $SO(2,1)$ element $L$)
and translate (over a three-vector $t$) to a frame where 
the particle is at rest in the origin, perform the closed loop to pick
up the rotation and translation due to its mass and spin, and then
transform back to his/her original restframe. This means that the
holonomy $U$ corresponding to such a particle is given by a conjugation
of $(\Lm_{\mu}, a_s)$ with the Poincar\'e element $(L,t)$. 
Using the definition of the adjoint action in the Lie algebra as given
in Eq.(\ref{eq:adj}) we find
\be
U = (\Lm, a) = (L,t)\,(\Lm_{\mu}, a_s)\,
   (L,t)^{-1} = \exp \left({\rm Ad}_{(L,t)}\,(\mu\cJ_0 +
    s\cP_o) \right) .
\label{eq:iso21holonomy}
\ee
It is not hard to see that in the presence of more than one particle
a closed loop around several particles will give the product of the
encircled holonomies, in which we recognise the fusion property as
described in subsection \ref{ssclasfusion}.

\vspace*{.2cm}

Let us return to the case of pure gravity, no particles present. An
important observation is that the Einstein--Hilbert action in three  
dimensions can be rewritten as the action of a Chern--Simons theory for
an $ISO(2,1)$ gauge field \cite{AchucTow} \cite{Witgrav}. To that aim
one uses the 
vielbein formalism for curved manifolds, so one introduces the dreibein
$e^{a}_{\mu}$ and the associated spin connection
${\om^{ab}}_{\mu}$. They obey the well known relation 
\be
g_{\mu\nu}(x)={e^a}_{\mu}(x) {e^b}_{\nu}(x)\, \eta_{ab},
\label{eq:gnee}
\ee
and the zero torsion condition
\be
T^a_{\mu\nu} = \partial_{\mu}{e^a}_{\nu} - \partial_{\nu}{e^a}_{\mu} +
{\om^a}_{b\mu} {e^b}_{\nu}-{\om^a}_{b\nu} {e^b}_{\mu} =0.
\ee
With the definition ${\om^a}_{\nu} = \half {\ep^a}_{bc}
{\om^{bc}}_{\mu}$ the Einstein--Hilbert action of
Eq.({\ref{eq:EinsteinHilbert}) can be written as
\be
S = \int d^3x\, \ep^{\mu\nu\rho}  e_{a\rho}\left(
\partial_{\mu}{\om^a}_{\nu} - \partial_{\nu}{\om^a}_{\mu} +
{\ep^a}_{bc}\left( {\om^b}_{\mu} {\om^c}_{\nu} -  {\om^b}_{\nu}
{\om^c}_{\mu} \right)\right).
\label{eq:EHineandomega}
\ee
Consider the $ISO(2,1)$ gauge field
\be
A_{\mu}= {e^a}_{\mu} \cP_{a}+ {\om^a}_{\mu} \cJ_{a},
\label{eq:eomgaugefield}
\ee
with $\cJ_a$ and $\cP_a$ the generators of the three-dimensional
Poincar\'e group $ISO(2,1)$, 
\be
[\cJ_a, \cJ_b]=\ep_{abc} \cJ^c,\quad [\cJ_a,\cP_b] = \ep_{abc}\cP^c,
\qquad [\cP_a,\cP_b] = 0,
\label{eq:comreliso21}
\ee
and the following inner product on the algebra
\be
\langle\cJ_a, \cP_b\rangle = \dl_{ab},\quad 
\langle\cP_a, \cP_b\rangle = \langle\cJ_a, \cJ_b\rangle = 0.
\ee
Then Eq.(\ref{eq:EHineandomega}) is nothing but the Chern--Simons
action as given in Eq.(\ref{eq:CSaction}) for the gauge field of
Eq.(\ref{eq:eomgaugefield}).
We have obtained the same theory as in section \ref{s2},
however, now for the inhomogeneous group $ISO(2,1)$, which is reflected
in the fact that the contraction of the flat indices $a,b,c,...$ goes
via the Minkowski metric $\eta_{ab}$. 

Inclusion of particles in the theory can be done in a similar fashion
as in section \ref{s2a}, that is, by allowing for curvature and torsion
at the positions of the particles, just as in Eq.(\ref{eq:nonvancurv}).
The resulting holonomy degrees of freedom are exactly the same as given
in Eq.(\ref{eq:hol}), but now take values in the group $ISO(2,1)$.
This coincides with the holonomy of Eq.(\ref{eq:iso21holonomy}), since
\be
\exp ({\rm Ad}_{(L,t)}(\mu\cJ_0 + s\cP_o)) = e^{p^a\cJ_a +
   j^a\cP_a},
\label{eq:iso21holtwo}
\ee
with $p^a = \Lm^a_0 p^0$ and $j^a = \Lm^a_0 j^0 + {\ep^a}_{bc} t^b
(\Lm p)^c$, compare Eq.(\ref{eq:coadaction}). In ordinary CS 
theory these $p^a$ and $j^a$ are internal degrees of freedom, taking
values in the dual of the Lie algebra. For 2+1 gravity, however, they
denote the real, external (space time) momentum and angular momentum of a
particle in the Minkowski frame at the position of the particle. 
%and there is no such internal space associated to a particle
%as used in the coadjoin orbit construction of subsection
%\ref{sssingleclas}. 
We return to this crucial point at the end of this section.

\vspace*{.2cm}

For a single particle the total phase space consists of the conjugacy
class labeled by $\mu$ and $s$ in ${\cal M}=ISO(2,1)$, which has a natural
cotangent bundle structure, just as in Eqs.(\ref{eq:multiphsp}) and
(\ref{eq:configspace}). This allows for a straightforward choice of
polarisation of the total phase space, and since the homogeneous part of
an $ISO(2,1)$ holonomy corresponds to the three-momentum $p^a$ of the
particle, it follows directly that upon
quantisation the Hilbert space consists of wave functions of the
momentum. For a given fixed mass $\mu$ this momentum $p^a$ lies on the
hyperboloid $(p^0)^2 - (p^i)^2 = \mu^2$, with $0<\mu<2\pi$ as it
should, and where $p^i$ denotes the spatial 
momentum. Note that, as in the case of $\iso$, the elements of
the configuration space, which are holonomies depending on
the {\sl momenta}, are noncommuting. Also the generators $j^a$ of
`translations' in configuration space are noncommuting, compare
Eq.(\ref{eq:pbfuncs}). 
 
Mathematically spoken we can proceed with quantisation just as we did
in section \ref{squant}. This means that for the case of 2+1 gravity the
Hilbert space decomposes into irreducible unitary 
representation of ${\cal D}(SO(2,1))$, the quantum double of
$SO(2,1)$. This in contrast to a decomposition into irreps of
$ISO(2,1)$, although the important difference mainly shows up in
the interacting multi-particle systems. The sectors of the theory are
labeled by conjugacy classes in $SO(2,1)$ and irreps of their
centralisers. Due to the locally compact structure of $SO(2,1)$ there
are more than just two different types of sectors, as was the case for
$\DS$ where we had the generic and the special representations
corresponding to $0<r<2\pi$ and $r=0, 2\pi$ respectively. Recall that
$SO(2,1)$ is isomorphic to $SL(2,\dubbelR)$ and to $SU(1,1)$. The
classification of 
the irreps of ${\cal D}(SL(2,\dubbelR))$ has been given in \cite{KM}. Of
the ten types of conjugacy classes described there, we associate the
first type to the massive particles in 2+1 gravity. This conjugacy class
is given by $C_{\mu} =\{ x g_{\mu} \xinv \,|\, x \in SO(2,1)\}$,
with $g_{\mu}$ corresponding to the diagonal element as defined in
Eq.(\ref{eq:eulermatrices}), for $0<\mu<2\pi$. Note that this
representation-theoretical 
restriction for the values of $\mu$ agrees with the physically
allowed values for the mass. The centraliser of such conjugacy classes
in $SO(2,1)$ has irreps labeled by $s \in \half\dubbelZ$. 
The corresponding Hilbert space $V^{\mu}_s$ is given by
\be
V^{\mu}_s := L^2_s(SO(2,1)) = \{\phi:SO(2,1)\to \CC\;|\;\; \phi(x h) = 
s(\hinv) \phi(x), \quad \forall h\in U(1)\},
\label{eq:so21hilbsp}
\ee
which equals the space of $V_s$ sections over the orbit $C_{\mu} \simeq
SO(2,1)/U(1)$, which as a space equals the two dimensional hyperboloid
$H^2$. We emphasise again that the $x$ here denote the {\sl direction}
of the momentum $p^a$, and have nothing to do with spatial
coordinates. From harmonic analysis on $H^2$
\cite{Helgasson} it is known that the 
Plancherel decomposition of $L^2(SL(2,\dubbelR)/SO(2))$ only contains
the continuous irreps of $SL(2,\dubbelR)$, also called the first and
second fundamental series \cite{Vil}. More precisely, a function on
$SL(2,\dubbelR)/SO(2) \simeq SO(2,1)/SO(2) \simeq SU(1,1)/U(1)$ can be
decomposed in a subset of the matrix elements of the continuous
irreducible unitary representations of $SO(2,1)$. For the case of a
$V_s$ section over $SO(2,1)/SO(2)$ this subset consists of matrix
elements of either the first or the second fundamental series, depending
whether the fixed right hand side label $s$ is integer or half-integer,
respectively. We introduce $\bar{s} = s\,\mbox{mod}\dubbelZ$, which can
either be $0$ or $\half$. An arbitrary element of $V^{\mu}_s$ can be
written as 
\be
{}^{\mu}\!\phi_s (x) = \sum_{\ep=0,\half} \dl_{\ep,\bar{s}}\,
\int_0^{\infty} \sum_{m'=m+\ep}\, c^{\ep}_m(\lm) \,\, \lm\,
\tanh\pi(\lm+i\ep)\, 
\,\goth{D}^{-\half+i\lm}_{m's}(x)\ d\lm, \qquad x\in SU(1,1)
\label{eq:so21phidecomp}
\ee
which should be compared to the case of $\DS$ in
Eq.(\ref{eq:su2phidecomp}). The integration in
Eq.(\ref{eq:so21phidecomp}) uses the Plancherel measure
$\rho_{\ep}(\lm)$, which normalises the 
functions $T^{\lm}_{ms}:= \goth{D}^{-\half+i\lm}_{ms}$ as follows
\be
\langle \, T^{\lm}_{ms}\,|\,T^{\lm'}_{ms}\,\rangle =
\frac{1}{\rho_{\bar{s}}(\lm)} \dl(\lm - \lm') = (\pi\lm\, \tanh
\pi(\lm+i\bar{s}) )^{-1}\,\dl(\lm - \lm').
\ee
So the set 
\be
\{\goth{D}^{-\half+i\lm}_{ms}\,|
\,s\,\mbox{fixed} ,\, \lm\in\dubbelR^+, m\in (\dubbelZ+\bar{s})\}
\label{eq:Vmusbasis}
\ee
forms a basis for $V^{\mu}_s$. Just as the Wigner functions $D^j_{mn}$
are related to the Jacobi polynomials $P^j_{mn}$ as given in
Eq.(\ref{eq:djmn}), the functions $\goth{D}^{-\half+i\lm}_{ms}(x)$ are
related to the so-called Jacobi functions $\goth{P}^j_{mn}$. For more
details we refer to \cite{Vil}. 

\vspace*{.2cm}

Summarising, due to the fact that 2+1 gravity with point particles can
be written as the CS theory of an inhomogeneous group, namely
$ISO(2,1)$, we find that its quantum theory is governed by the quantum
double of the homogeneous part of that group, ${\cal D}(SO(2,1))$. 
For a single particle with fixed mass
$0<\mu<2\pi$ and spin $s\in \half\dubbelZ$ we have found the total
Hilbert space $V^{\mu}_s$ of Eq.(\ref{eq:so21hilbsp}), i.e.\ the space
of wave functions for the metric and particle together, under the
constraints that the curvature and torsion of the gauge field are equal
to the momentum and angular momentum, respectively. The wave functions 
depend on the energy and momentum, and transform covariantly under right
action with $N_{\mu}=U(1)$. Due to the properties of the
comultiplication of the quantum double the fusion properties of 
multi-particle systems are automatically reflected on the quantum level,
analogously to the case of $\DS$ CS theory, which is described in 
detail in section \ref{s5} of the current article. Thus, a consistent
construction of a multi-particle Hilbert space in 2+1 gravity is (at
least partially) solved by identifying the 
underlying quantum double structure. This is supported by the fact that
the quantum double provides us with another important tool, namely the
$R$-element, which assures that the Hilbert space also forms a
representation of the braid group $B_n(\Sg)$ for $n$ distinguishable
particles. In principle this allows us to compute differential
cross-sections of scattering processes with massive, spinning particles
in 2+1 gravity. However, here one faces the interesting problem of
reconstructing the physical space time from the 
Hilbert space which arises upon canonical quantisation of the
multi-particle phase space ${\cal M} =
\mbox{Hom}(\pi_1(\Sg,*);ISO(2,1))/\sim$. This problem has been recognised
and addressed by other authors as well \cite{GriNar} \cite{VazWit}, but
they impose different matching conditions on the space time
coordinates. We will return to these issues in a forthcoming paper
\cite{BMgrav}.   

\section{Conclusions}\label{s7}
\setcounter{equation}{0}
We have shown that the representation theory of $\DS$, the quantum double of
$SU(2)$, can be used to give a quantised description of purely
topological interacting point particles coupled to an $\iso$ CS gauge
field in 2+1 dimensions. We 
discussed the classical properties of the system, and used
the explicit model of the $\iso$ CS theory to get 
an explicit parametrisation of the physical phase space. Due to the
fact that (classical) observables are only measured via
noncontractable loops the phase space has a multiplicative structure,
which must be maintained on a quantum level. This meant that we had
to quantise the particles {\sl plus} their interactions in one go,
instead of quantising single, free particles and describing their
interaction via an interaction term in the Hamiltonian. Indeed, due to the
topological nature of the interactions, it turned out to be 
possible to directly incorporate the interactions on the level of the
Hilbert space. In section \ref{squant} we
discussed the properties of the quantum system for single and
multi-particle cases, and the two types of interactions, `fusion' and 
`braiding'. In section \ref{s3} we made a side step to pure
mathematics, and described the quantum double $\DG$ of a (compact) group
$G$. Subsequently we gave its irreducible unitary representations, and
worked them out for the case $G=SU(2)$. Simply by 
comparing the labels of the irreps, their degrees of freedom, and their
properties under fusion and braiding, with the quantum numbers,
configuration space, and interaction properties of the wave functions
of the quantised system, we showed that the representation theory of
${\cal D}(SU(2))$ precisely covers the quantum theory of our system.

The nontrivial comultiplication, which is part of the Hopf algebra
structure of the quantum double, implies that the action 
on tensor product states is different from the tensor product action of
groups, and therefore that the decomposition in irreducible unitary
components (the Clebsch--Gordan series) becomes different and much more 
involved than for the case of ordinary groups. In section \ref{s5}
we worked it out in detail for the 
case of $G=SU(2)$. Although it is a rather technical aspect, it is
interesting to note that the derivation of the Clebsch--Gordan
coefficients for ${\cal D}(SU(2))$ requires solution of the problem of
combining Wigner functions in {\sl different} points of $SU(2)$. 
Next we studied the braiding of two particles. First we
described how the $R$-element of the quantum double describes the effect of
braiding, then we used the nonabelian generalisation of
the Aharonov--Bohm formula, Eq.(\ref{eq:crossec}), to calculate the
differential cross-section for various scattering processes. 
We concluded by pointing out the quantum double structure ${\cal
D}(SO(2,1))$ underlying the quantum theory of 2+1 gravity, as a preview
to \cite{BMgrav}. 

Finally, the connection with the `ordinary' $q$-deformed structure for CS
theories with homogeneous, compact gauge group deserves further
analysis. Related to this for the inhomogeneous case is the
identification of the `algebra of observables', 
corresponding to the functions on classical phase space, and leading to
the quantum double structure (or $q$-deformed structure for compact CS
theories) of the quantum theory.

%In fact, we have only concentrated on one aspect of
%quantisation, that is, we have constructed a Hilbert space of wave
%functions on the classical configuration space (with all the correct
%interactions), but have not yet given the `algebra of observables'
%corresponding to the functions on classical phase space. 
%This will be done in a follow-up of this paper. 
%
%
%
\appendix
\begin{appendix}
\section{The quantum double }
We summarise the construction of the quantum double for a locally compact
group as given in \cite{KM}. More generally the quantum double of a Hopf
algebra has been introduced by Drinfel'd in \cite{Drin}. Quantum doubles are
important examples of quasitriangular (quasi) Hopf algebras, and are 
well-studied in mathematics, see also the introduction in \cite{KBM}.
Readers who are not familiar with Hopf algebras we refer to 
\cite{CharPres} and \cite{Majbook}. A much more compact introduction to
quantum groups is given in \cite{Tjark}.
\begin{defn}
Let ${\cal A}$ be a Hopf algebra and ${\cal A}^0$ 
the dual Hopf algebra to ${\cal A}$ with the opposite comultiplication. 
Then $\DA$ is the unique quasitriangular Hopf algebra with universal
R-matrix $R\in \DA\otimes \DA$ such that
\begin{itemize}
\item[i.]  As a vector space, $\DA = {\cal A}\otimes{\cal A}^0$.
\item[ii.] ${\cal A} = {\cal A}\otimes 1$ and ${\cal A}^0 = 1 \otimes
{\cal A}^0$ are Hopf subalgebras of $\DA$.
\item[iii.] The mapping $x \otimes \xi \mapsto x\xi : {\cal A}\otimes
{\cal A}^0 \to \DA$ is an isomorphism of vector spaces. Here $x\xi$ is
short notation for the product $(x\otimes 1)(1\otimes\xi)$. 
\item[iv.] Let $(e_i)_{i\in I}$ be a basis of ${\cal A}$ and 
$(e^i)_{i\in I}$ the dual basis of ${\cal A}^0$. Then 
\be
R = \sum_{i\in I} (e_i \otimes 1)\otimes
(1\otimes e^i),
\ee
independent of the choice of the basis. 
\end{itemize}
\end{defn}
For a finite group $G$ we take ${\cal A} := C(G)$, the space of all complex
valued functions on $G$, which becomes a Hopf algebra under pointwise 
multiplication:
\be
(f_1 \cdot f_2) (x) = f_1(x) f_2(x), \qquad x \in G
\ee
with comultiplication
\be
(\Dl f)(x,y) := f(xy), \qquad x,y \in G
\label{eq:comfuncalg}
\ee
and with antipode
\be
(Sf)(x) := f(\xinv), \qquad x \in G.
\ee
The dual of $C(G)$ is the group algebra $\CC[G]$ with as multiplication the
ordinary group multiplication, and with comultiplication
\be
\Dl (x) = x \otimes x, \qquad x\in G.
\ee
The antipode is the inverse in the group. The pairing between $C(G)$ and $\CC[G]$
is given by
\be
\langle f,x \rangle = f(x), \qquad f\in C(G), x \in G.
\ee
\vspace*{.2cm}

For the quantum double of $C(G)$ we now write
\be
{\cal D}(G) := {\cal D}(C(G)) = C(G) \otimes \CC[G] \simeq C(G,\CC[G]).
\ee
Thus $f\otimes g \in C(G) \otimes \CC[G]$ can be considered as a mapping 
from  $G \to \CC[G]$ via
\be
(f\otimes g):z \mapsto f(z) g .
\ee
Also ${\cal D}(G) \otimes {\cal D}(G) \simeq C(G\times G, \CC[G]\otimes 
\CC[G])$. The full quasi triangular Hopf algebra structure of $\DG$ is given by
\begin{itemize}
\item  Multiplication
\be
(f_1 \otimes g_1)(f_2 \otimes g_2) = f_1(.) f_2(\ginv_1 .\,g_1) \otimes g_1 g_2
\quad:\quad  z \mapsto f_1(z) f_2(\ginv zg)\: g_1 g_2. 
\label{eq:muldub}
\ee
Especially
\be
(1 \otimes g)(f \otimes e) = f(\ginv .\,g)\otimes g, \qquad (e\,\,
\mbox{unit in}\, G). 
\ee
\item Comultiplication
\be
(\Dl (f\otimes g))(y,z) = f(yz)\, g\otimes g, \qquad (g,y,z \in G, f\in C(G))
\label{eq:comdub}
\ee
\item Unit
\be
1\otimes e \quad:\quad z \mapsto e. 
\label{eq:undub}
\ee
\item Co-unit
\be
\ep(f\otimes x) = f(e).
\ee
\item Antipode
\be
S(f\otimes g) = f(g (.)^{-1} \ginv)\otimes \ginv \quad:\quad z \mapsto 
f(g \zinv \ginv)\:\ginv.
\ee
\item $R$-element
\be
R = \sum_{g \in G} (\dl_g \otimes e) \otimes (1 \otimes g)\quad:\quad (x,y) 
\mapsto e\otimes x,\qquad(x,y)\in G\times G 
\label{eq:rmatrix}
\ee
where $\dl_g$ is the Kronecker delta on $g \in G$, and thus a (basis) element
of $C(G)$. 
\end{itemize} 
As explained in \cite{KM} this construction cannot be uniquely generalised
to the case of continuous $G$, even not for compact $G$. To avoid the 
difficulties the construction can be reformulated with aid of the following
linear bijection:
\be
\DG = C(G) \otimes \CC[G] \Longleftrightarrow C(G\times G),
\label{eq:biject}
\ee
for which we can write
\bea
\fog &\mapsto& \left((x,y) \mapsto f(x) \dl_g(y)\right) \nn\\
\sum_{z\in G} F(.\, ,z)\otimes z & \leftarrow& F
\eea
\end{appendix}
\subsection*{Acknowledgements}
We thank Prof.\ T.H.\ Koornwinder, Prof.\ H.L.\ Verlinde, and Dr B.\
Schroers for valuable discussions.  The second author is supported by
the Dutch Science Foundation FOM/NWO.

\end{document}